\documentclass[usenatbib]{mn2e}
\usepackage{epsfig,rotating,natbib,amssymb,amsmath,times}

\title[Magnetic fields in star cluster formation]{The effect of magnetic fields on star cluster formation}

\author[Price \& Bate]{Daniel J. Price and Matthew R. Bate \\
School of Physics, University of Exeter, Stocker Rd, Exeter EX4 4QL, UK \\
}
\date{Submitted: Revised: Accepted: }
\pagerange{\pageref{firstpage}--\pageref{lastpage}} \pubyear{2007}

\begin{document}
\label{firstpage}
\bibliographystyle{mn2e}
\maketitle

\begin{abstract}
 We examine the effect of magnetic fields on star cluster formation by performing simulations following the self-gravitating collapse of a turbulent molecular cloud to form stars in ideal MHD. The collapse of the cloud is computed for global mass-to-flux ratios of $\infty$, $20$, $10$, $5$ and $3$, that is using both weak and strong magnetic fields. Whilst even at very low strengths the magnetic field is able to significantly influence the star formation process, for magnetic fields with plasma $\beta < 1$ the results are substantially different to the hydrodynamic case. In these cases we find large-scale magnetically-supported voids imprinted in the cloud structure; anisotropic turbulent motions and column density structure aligned with the magnetic field lines, both of which have recently been observed in the Taurus molecular cloud. We also find strongly suppressed accretion in the magnetised runs, leading to up to a 75\% reduction in the amount of mass converted into stars over the course of the calculations and a more quiescent mode of star formation. There is also some indication that the relative formation efficiency of brown dwarfs is lower in the strongly magnetised runs due to the reduction in the importance of protostellar ejections.
\end{abstract}


\section{Introduction}
 Understanding how stars like the Sun form is one of the key questions central to our understanding of the Universe we live in. Whilst we have come a long way in this understanding since the pioneering work of Jeans, many of the fundamental questions such as the rate, distribution and efficiency of star formation remain either unknown or are the subject of vigourous debate. One of the key areas of uncertainty is whether or not star formation is a rapid \citep{mk04,hbb01,elmegreen07} or slow \citep{sla87,tkm06} process, central to which is the relative importance of magnetic fields to the star formation process.

 Whether or not star formation is rapid or slow, the fact remains that molecular clouds are observed to contain magnetic fields of sufficient strengths that they cannot be ignored in any complete theory of how stars form from such clouds \citep[e.g.][]{crutcher99,bourkeetal01,hc05}. Furthermore molecular clouds are observed to contain supersonic turbulent motions \citep{larson81}, so the interaction of turbulence and magnetic fields is critical to our understanding of the star formation process. This interaction has been the subject of a number of studies which have shown that magnetic fields are \emph{not} effective in preventing the rapid dissipation of supersonic turbulence in the absence of continued driving \citep{sog98,maclowea98,vsetal00}, though the presence of magnetic fields does change the dynamics of the turbulence \citep{pn02,padoanetal07}. However, to date, there has been only a handful of simulations which have attempted to follow the self-gravitating collapse of a turbulent cloud in the presence of a magnetic field \citep[e.g.][]{lietal04,ln06,vsetal05,tp07}.

 The ability of magnetic fields to provide support against gravitational instability is determined, for an enclosed region of gas threaded by a magnetic field, by the ratio of the mass contained within the region to the magnetic flux passing through the surface, ie. the \emph{mass-to-flux} ratio, which for a spherical cloud is given by
\begin{equation}
\frac{M}{\Phi} \equiv \frac{M}{4\pi R^{2} B_{0}},
\label{eq:masstoflux}
\end{equation} 
where $M$ is the mass contained within the cloud volume, $\Phi$ is the magnetic flux threading the cloud surface at radius $R$ assuming a uniform magnetic field $B_{0}$. The critical value of $M/\Phi$ below which a cloud will be supported against gravitational collapse is given by \citep[e.g.][]{ms76,mestel99,mk04}
\begin{equation}
\left(\frac{M}{\Phi}\right)_{\rm crit} = \frac{2 c_{1}}{3} \sqrt{\frac{5}{\pi G \mu_{0} }},
\label{eq:mphicrit}
\end{equation}
where $G$ and $\mu_{0}$ are the gravitational constant and the permeability of free space respectively and $c_{1}$ is a constant determined numerically by \citet{ms76} to be $c_{1}\approx 0.53$. 
Star forming cores with mass-to-flux ratios less than unity are stable against collapse (``subcritical'') and conversely, cores with mass-to-flux ratios greater than unity (termed ``supercritical'') will collapse on the free-fall timescale. Throughout this paper we use the mass-to-flux ratio, defined in terms of the critical value, to quantify the magnetic support of a molecular cloud against collapse and we perform simulations using initially supercritical clouds (though to varying degrees). Giant molecular clouds are generally thought to have magnetically subcritical envelopes but to be supercritical in their inner parts \citet{cm95,ccw05}, a picture which is largely confirmed by observational results on both large scales \citep{mckee89,mckeeetal93} and in dense cores \citep{crutcher99,bourkeetal01}. 

 \citet{lietal04} examined the role of magnetic fields in self-gravitating core formation within a turbulent molecular cloud in a periodic box. They found that cores formed within a supercritical cloud were also locally supercritical by at least an order of magnitude, indicating that a globally supercritical magnetic field does \emph{not} evolve to produce locally magnetically subcritical cores, contrary to some earlier expectations \citep{ms56}. Furthermore, even with supercritical cores they found a central magnetic field strength in cores $B \propto \rho^{1/2}$, similar to observations \citep{crutcher99} which has often been used as an argument for magnetic support in molecular cloud cores. They also found strong interaction between cores and rotationally supported discs. \citet{tp07} also performed simulations of self-gravitating collapse in the presence of magnetic fields and found that the observed near-critical cores could form naturally from a globally highly supercritical cloud, and that these cores generally have the same gas-to-magnetic pressure ratio, $\beta$ as the mean $\beta$ in the global cloud.
 
 Given that magnetic fields always act to oppose gravitational collapse, it has often been suggested that they may play the dominant role in regulating the star formation efficiency in molecular clouds \citep[e.g.][]{kt07}, though there are also several other good candidates for doing so, including turbulence-inhibited collapse \citep[e.g.][]{tp04,lietal04,vsetal05}, feedback from jets and outflows \citep{nl05,ln06}, the dispersal of initially unbound clouds \citep{clarketal05} or radiation feedback from the stars themselves. Secondly magnetic fields are often invoked to solve the `angular momentum problem' in star formation via the magnetic braking of star forming cores. Recent simulations by \citet{pb07} have shown that magnetic fields can have a dramatic effect on circumstellar disc formation and on fragmentation to form binary systems (these results have since been confirmed by \citealt{ht08}). \citet{tp07} looked at the effect of the magnetic fields on the star formation efficiency in their simulations, though no clear trend was apparent, in part due to limitations of the numerical model.

 Despite the apparent importance of magnetic fields to the star formation problem, it was therefore somewhat surprising that the purely hydrodynamic calculations of \citet{bbb03} (hereafter BBB03) produced largely the `right answer' in terms of being able to reproduce observed the stellar initial mass function (albeit at low number statistics) as well as several other observational characteristics such as the frequency of binary stars and stellar velocity dispersions. In this and subsequent calculations the initial mass function is built up due to the competition between dynamically interacting protostars in order to accrete from the global cloud \citep{bb06}. Thus, low mass stars are simply those which have been quickly ejected from multiple systems and thus have only a short accretion history \citep{bbb02a,bb05}, whereas higher mass stars are those which form and remain at the bottom of deep potential wells and build up their mass through accretion over time \citep{bonnelletal97}. However subsequent calculations (Bate 2008, in prep) have established that purely hydrodynamic calculations produce an excess of brown dwarfs relative to observations.
 
  Thus it is crucial to extend these types of calculations, with the hope of resolving some of the above issues, by adding the two major pieces of missing physics -- magnetic fields and the effect of radiation transport. This paper presents our first attempt to address the former in large-scale simulations. Whilst in a sense the two are complementary, since we expect the magnetic field to have an effect primarily on larger scales (given the strong physical diffusion of the magnetic field on smaller scales), whilst radiation might be expected to influence the smaller scale dynamics (ie. fragmentation), it is clear that it is imperative to incorporate both pieces of physics into these types of calculations.

 The paper presents our first investigation of how magnetic fields change the picture of star cluster formation painted by BBB03. The numerical method is discussed in \S\ref{sec:numerics} and the initial conditions for the simulations are discussed in \S\ref{sec:initconds}.  Results are presented in \S\ref{sec:results} and discussed in \S\ref{sec:discussion}.

\section{Numerical method}
\label{sec:numerics}

\subsection{Hydrodynamics}
 We solve the equations of self-gravitating (magneto-) hydrodynamics using the Smoothed Particle Hydrodynamics (SPH) method (for recent reviews of SPH see \citealt{monaghan05,price04}). Fluid quantities and their derivatives in SPH are evaluated on a set of moving particles which follow the fluid motion. The long range gravitational force is calculated efficiently using a binary tree algorithm originally written by \citet{benzetal90}, although substantial modifications have been made to the code since, both in terms of efficiency and as improvements to the basic algorithms. Individual timesteps were added by \citet{bate95} and sink particles (discussed below) were implemented by \citet*{bbp95}. The code at this stage was used for the original BBB03 calculations.
 
  More recently (that is, post BBB03), the hydrodynamics in the code has been thoroughly updated with state-of-the-art SPH algorithms, most notably by adopting the energy and entropy-conserving variable smoothing length algorithms developed by \citet{sh02}, \citet{monaghan02} and \citet{pm04b} and by the introduction of additional physics in the form of magnetic fields \citep{pb07} and radiative transfer using the flux-limited diffusion approximation \citep{wb05} (although we do not include radiative transfer in this paper). In the variable smoothing length formulation the density for each particle is calculated according to
\begin{equation}
\rho_{i} = \sum_{j} m_{j} W_{ij} (h_{i}) \label{eq:dens}
\end{equation}
where the smoothing length $h_{i}$ is itself a function of the density in the form
\begin{equation}
h = \eta \left( \frac{m}{\rho} \right)^{1/3},
\end{equation}
where $\eta$ is a parameter determining the approximate neighbour number (here we choose $\eta = 1.2$ corresponding to approximately 60 SPH neighbours). Thus the density summation (\ref{eq:dens}) becomes a non-linear equation for both $h$ and $\rho$ which we solve iteratively as described in \citet{pm07}. 

 Short range gravitational forces (ie. between particles lying within each others smoothing spheres) are softened using the SPH kernel with a softening length which is set equal to the SPH smoothing length for that particle. We formulate the force softening using the formalism presented recently by \citet{pm07} which ensures that momentum and energy are \emph{both} conserved even though the softening length is a spatially variable quantity.

\subsection{Magnetohydrodynamics}
 The magnetohydrodynamics in the code is based on the recent development of MHD in SPH by \citet{pm04a,pm04b} and \citet{pm05}. For the calculations presented here, as in \citet{pb07}, we use the `Euler potentials' formulation for the magnetic field such that the divergence constraint is satisfied by construction. For more details of the Euler potentials formulation we refer the reader to \citet{pb07} and to a complete description of a code similar (though not identical) to that used here given in \citet{rp07}.
 
  Use of the Euler potentials for the magnetic field evolution is a slightly more limited formulation of MHD than that presented in \citet{pm05} in that certain types of initial field geometry cannot be represented in such a formulation (see discussion in \citealt{rp07}). Whilst this does not present immediate difficulties for the simulations presented in this paper (starting with a uniform field geometry), we would, for example, not expect dynamo processes to be well captured in the Euler potentials formulation because of the helicity constraint. 

  Shocks in both hydrodynamics and MHD are captured via dissipative terms corresponding to an artificial viscosity \citep{monaghan97} and for MHD, resistivity \citep{pm04a,pm05} with controlling parameters which are individual for each particle and evolve with time (thus reducing dissipation away from shocks) as described in \citet{pm05} based on the original formulation of \citet{mm97}. The formulation of artificial resistivity in the Euler potentials' evolution is described in \citet{pb07} and \citet{rp07}. For reference, as in \citet{pb07} the magnetic force is formulated using the `Morris formulation' described in \citet{pm05} which is both stable in the regime where the magnetic pressure exceeds the gas pressure whilst conserving momentum sufficiently for the accurate simulation of shocks. 
 
  A major limitation to the simulations presented in this preliminary work is that, at the resolution of the original BBB03 calculation (which was determined by the criterion that all of the hydrodynamic fragmentation was resolved in the simulation), the artificial resistivity plays a dominant role in the evolution of the magnetic field on small scales (that is, during the actual collapse to form stars). Whilst it may be argued that ideal MHD is also a poor approximation for real molecular clouds, it is a wholly undesirable situation to have numerical dissipation in place of physical dissipation effects. Thus, for example, we are not able at this resolution to confidently assert that we have accurately captured the influence of the magnetic field on the fragmentation of individual cores (e.g. as in \citealt{pb07}). Instead we limit ourselves to a discussion of the influence of the magnetic field on the large scale structure of the cloud and details such as the initial mass function produced in the MHD runs should be taken with the appropriate degree of caution. Future calculations will be performed at a much higher resolution in order to follow the structure of the magnetic field further into the collapse.

\subsection{Equation of state}
\label{sec:eos}
 The effects of radiative transfer are approximated by adopting an equation of state of the form:
\begin{equation}
P = K \rho^{\gamma}.
\end{equation}
where the polytropic exponent $\gamma$ is given by
\begin{eqnarray}
\gamma = 1,  & & \rho \le 10^{-13} {\rm g\phantom{l}cm}^{-3}, \nonumber \\
\gamma = 7/5, & &  \rho > 10^{-13} {\rm g\phantom{l}cm}^{-3}.
\label{eq:eos}
\end{eqnarray}
The equation of state is isothermal at low densities ($< 10^{-13}$g cm$^{-3}$) where heating and cooling in molecular clouds balance. At higher densities the equation of state becomes barytropic with the polytropic exponent chosen to match the results of one dimensional (spherically symmetric) calculations which include the full effects of radiative transfer \citep{mi00} (see BBB03).

The net effect of the above equation of state is that collapse proceeds unhindered until the density reaches the critical density, at which point the gas begins to heat as it is compressed, providing thermal support which resists collapse. For hydrodynamics this critical density sets the minimum fragment mass from which objects subsequently accrete \citep{bate05}.

 We caution that the adoption of an equation of state of the form (\ref{eq:eos}) is based on conditions at the centre of a spherically symmetric prestellar core of $1M_{\odot}$. Thus this approximation may be expected to break down in regions where spherical symmetry is broken -- most notably this may be true for fragmentation occurring in discs. Also, the equation of state in the form (\ref{eq:eos}) only depends on the local gas density and thus does not account for the propagation of radiation which would increase the temperature in the material surrounding the protostars. Equation (\ref{eq:eos}) is also a rather crude parameterisation even of the 1D \citet{mi00} calculations, and the effect of changes to the assumed polytropic index has not been investigated. The effect of these assumptions on the cloud fragmentation is at present uncertain and will ultimately require calculations including a self-consistent treatment of radiative transfer. Fortunately such simulations, whilst not yet including magnetic fields, are starting to be performed \citep[e.g.][]{wb06} based on the radiative transfer algorithms developed by \citet*{wb05}.

\subsection{Sink Particles}
 Sink particles were introduced into SPH by \citet*{bbp95} to enable star formation calculations to be followed beyond the point at which stars form in order to study the subsequent global cloud dynamics rather than the internal dynamics of the stars themselves. In the calculations described in this paper, sink particles are allowed to form (that is, the SPH particle lying closest to the density maxima is converted into a sink particle) once the following conditions are satisfied:
\begin{enumerate}
\item the density exceeds $5.5\times 10^{-9}$g cm$^{-3}$ (the exact number is somewhat arbitrary -- a lower number means that some dynamics may be missed whilst a higher number means substantial slowdown of the code whilst trying to evolve material at extremely high densities)
\item a Jeans mass of material is contained within a kernel radius (ie. twice the smoothing length) of the particle
\item the material which will form the sink has a ratio of thermal to gravitational energy, $\alpha_{grav} < 0.5$ and $\alpha_{grav} + \beta_{grav} < 1.0$ (where $\beta_{grav}$ is the ratio of rotational energy to the magnitude of the gravitational energy). 
\item the divergence of velocity at the particle location is negative (ie. material is collapsing)
\item all these particles are being moved on the current timestep
\end{enumerate}

 One pitfall in rigourously enforcing the above criteria is that it is possible that material within a kernel radius, though part of a larger bound object, is not self-bound, unnecessarily blocking sink particle formation. A simple method of avoiding this, adopted in this paper, is to revert the equation of state to isothermal once a density of $10^{-11}$g cm$^{-3}$ has been reached in order to pass the $\alpha$ and $\beta$ tests and force the formation of a sink particle beyond this density. For all practical purposes this is almost identical to simply overriding tests ii) and iii) and inserting a sink particle at a density of $10^{-11}$g cm$^{-3}$ regardless. It is important to note that this artificial change in the equation of state has no effect on the fragmentation because the gas never fragments on these scales (ie. in none of the calculations do we get sink particles forming close to each other).
  
 Sink particles, once created, are subsequently allowed accrete all material which falls within a fixed accretion radius (set to 5AU in the calculations presented here) and is bound to the sink. Short range gravitational encounters between sink particles are softened using the cubic spline kernel where in these calculations we have used a fixed softening length of $4$AU for sink-sink interactions (note that the softening length for the self-gravitating gas is always set equal to the SPH smoothing length and thus varies according to the gas density).

 Note that in these calculations, as in previous works \citep{bbb03,bb05}, a sink particle is \emph{only} allowed to form once the gas has become optically thick to radiation and thus no further fragmentation is expected. Thus, provided the Jeans mass is sufficiently resolved, the calculations are expected to resolve \emph{all} of the fragmentation present (that is the initial mass function is expected to be complete at the low mass end apart from extremely hard binaries with orbital separations $<$ 5AU, if indeed fragmentation is possible at such scales).

\section{Initial conditions}
\label{sec:initconds}


\subsection{Cloud properties}
 The initial molecular cloud is set up identically to that described in BBB03: The cloud is initially spherical with a diameter of $0.375$ pc (77,400 AU) and contains a total of 50$M_{\odot}$ of material, uniformly distributed, giving an initial density of $\rho_{0} = 1.225 \times 10^{-19}$g cm$^{-3}$ ($n_{H_{2}} = 3.7\times 10^{4}$). The cloud free-fall time is $t_{ff} = \sqrt{3\pi/(32\rho_{0} G)} = 1.90 \times 10^{5}$ yrs. Thus in observational terms this corresponds to a small, relatively dense patch of a molecular cloud (although the early evolution whilst the gas remains isothermal is scale free). Particles are placed in a uniform random distribution cropped to the cloud radius and no particles are placed exterior to the cloud, resulting in a significant expansion of the outer layers as the collapse proceeds (this would be equivalent to the assumption of open boundary conditions in a grid-based simulation). The cloud is given an initial sound speed of $1.84 \times 10^{4}$ cm/s, which corresponds to a temperature of $\sim10$K (the exact temperature depends on the assumed value for the mean molecular weight). The resultant ratio of thermal to gravitational energy is $\alpha_{grav} = 0.073$.

 A total of 3.5 million SPH particles are used in each of the calculations (as in BBB03). As previously mentioned, the resolution for a cloud of this size was determined in BBB03 by the requirement that fragmentation should be resolved according to the \citet{bateburkert97} criterion. Given the recent improvements to the hydrodynamic algorithm alongside the incorporation of MHD, in one sense the purely hydrodynamic calculation presented here (see \S\ref{sec:results}) may be viewed as a repeat of the BBB03 calculation with a thoroughly updated SPH method, although there are slight differences in the initial conditions (discussed below) which mean that the hydrodynamic calculation presented here is also not completely identical to the original run.



\subsection{Turbulent velocity field}
 The cloud is imprinted with a turbulent velocity field as described in BBB03 with power spectrum $P(k)\propto k^{-4}$. Whilst the generated velocity field (produced on a grid) is similar to that used in the \citet{bbb03} calculations, a slight change in the way in which the velocity field was interpolated to the particles has been subsequently added to the code and this means that, whilst the hydrodynamic evolution is very similar to the \citet{bbb03} calculation, it is not identical. The initial velocity field is normalised such that the kinetic energy is initially set equal to the gravitational potential energy of the cloud, which gives an initial Root Mean Square (RMS) Mach number of 6.4 and an initial RMS velocity of $1.17 \times 10^{5}$ cm/s.

\begin{figure*}
\begin{center}
\includegraphics[width=\textwidth]{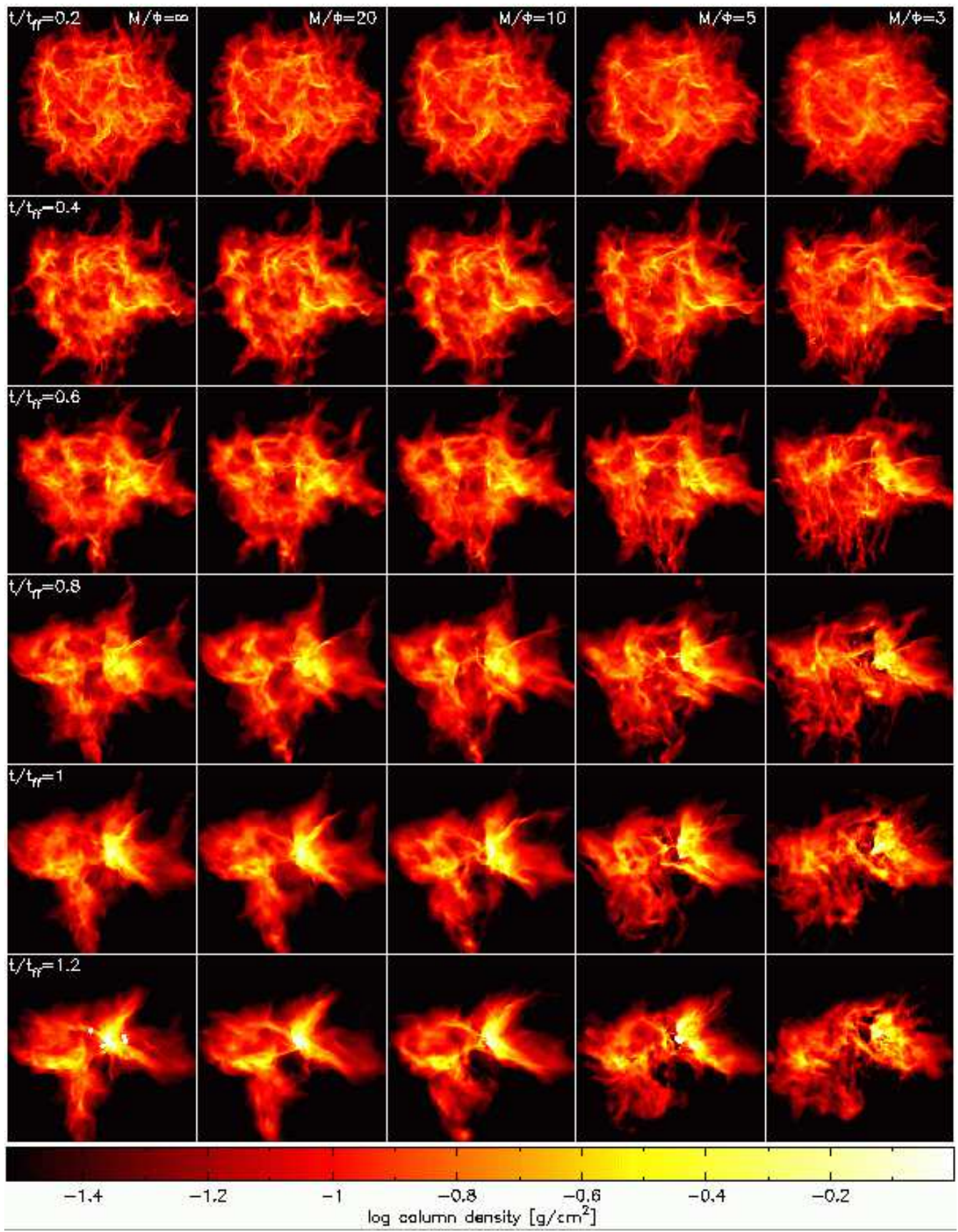}
\caption{Global cloud evolution, shown as column density in the cloud at intervals of $0.2$ cloud free-fall times (top to bottom) for the five runs of progressively increasing magnetic field strength (left to right), parametrised in terms of the mass-to-flux ratio of the cloud in units of the critical value. Thus $M/\Phi = \infty$ is a hydrodynamic evolution whilst the strongest field run is $M/\Phi = 3$ (that is, supercritical by a factor of 3). Note the large voids and vertical filamentary structure in the strongly magnetised runs.}
\label{fig:globalcloud}
\end{center}
\end{figure*}

\subsection{Magnetic fields}
\label{sec:magfields}
 We perform a sequence of calculations with an initially uniform magnetic field of progressively increasing strength threading the cloud. The strength is parameterised in terms of the mass-to-flux ratio expressed in units of the critical value, where we have performed runs using $M/\Phi = \infty$ (ie. no magnetic field), 20, 10, 5 and 3. All of the values are supercritical since given the absence of ambipolar diffusion in our calculations, subcritical clouds would not be expected to collapse. We have verified that this is indeed the case by also performing a calculation at a mass-to-flux ratio of unity (ie. critical) which, as expected, does not collapse to form stars (the cloud flattens along the direction of the magnetic field but undergoes a bounce and subsequent expansion rather than collapse).


 Given the cloud dimensions and the mass-to-flux ratio, corresponding physical magnetic field strengths can be determined for each of the runs according to
\begin{equation}
B_{0} = 194 \mu G \left(\frac{M}{\Phi} \right)^{-1} \left(\frac{M}{50 M_{\odot}}\right) \left(\frac{R}{0.1875 {\rm pc}}\right)^{-2},
\end{equation}
where $M/\Phi$ is the mass to flux ratio in units of the critical value. Thus a run with a critical mass-to-flux ratio would have $B_{0} = 194\mu G$ and for the runs with mass-to-flux ratios of $\infty, 20, 10, 5$ and $3$ the corresponding field strengths are given by $B_{0} = 0, 9.7, 19, 39$ and $65 \mu G$ respectively.

 The magnetic field may also be parametrised in terms of the plasma $\beta$, the ratio of gas to magnetic pressure, according to
\begin{equation}
\beta =  0.0276 \left(\frac{M}{\Phi} \right)^{2} \left(\frac{c_{s}}{18.4 {\rm km/s}}\right)^{2} \left(\frac{M}{50 M_{\odot}}\right)^{-1} \left(\frac{R}{0.1875 {\rm pc}}\right).
\end{equation} 
  The five runs presented here thus have initial $\beta$'s of $\infty$, $11$, $2.8$, $0.7$ and $0.25$ respectively. Note that the magnetic pressure is dominant over gas pressure in the cloud for mass-to-flux ratios $< 6$ which is the case for the two strongest-field runs. Indeed we find that these two runs shown far more significant differences compared to the weaker field and hydrodynamic runs.

The Alfv\'en speed in the initial cloud is given by
\begin{equation}
v_{A} =  1.57 \times 10^{5} {\rm cm/s} \left(\frac{M}{\Phi} \right)^{-1} \left(\frac{M}{50 M_{\odot}}\right)^{1/2} \left(\frac{R}{0.1875 {\rm pc}}\right)^{-1/2},
\end{equation}
giving $v_{A} = 0$, $7.8 \times 10^{3}$, $1.6 \times 10^{4}$, $3.1 \times 10^{4}$ and $5.2 \times 10^{4}$ cm/s for the five runs. Thus, the initial turbulent motions in the cloud are super-Alfv\'enic in all cases with Alfv\'enic Mach numbers of $\infty$, $15$, $7.3$, $3.8$ and $2.3$ respectively.

  The initial magnetic field is defined as a linear gradient in the Euler potentials on the particles. Since the gradient of the Euler potentials is computed exactly to linear order \citep[][]{pb07,rp07}, the field is thus uniform everywhere (including at the free boundary). As the calculation progresses the field is naturally carried by and thus anchored to a surrounding medium created by the expansion of the outer layers of the cloud (see above). This initial evolution of the field is discussed further in \S\ref{sec:fieldevol} and shown for each of the simulations in Figure~\ref{fig:fieldevol}.

 It is worth briefly discussing the validity of starting the calculation with an initially imposed uniform magnetic field, since clearly in reality there will be a mixture of random and ordered components in the field of varying magnitude. However, we also start with a uniform density cloud, so density structure and non-uniformity in the magnetic field are both generated self-consistently by the initially imposed turbulent velocity field (as opposed to starting with pre-existing density structure on which a magnetic field is imposed). An alternative approach which could be explored in future calculations might be to start with a turbulent box containing a magnetic field which has been artificially driven to a saturated state (in the absence of self-gravity), although even in this case it is not clear that this would correspond any better to reality, since molecular clouds are clearly not periodic structures and the sudden ``turning on'' of self-gravity is equally questionable. Starting with a uniform magnetic field does however provide a meaningful upper limit to the effect of the magnetic field on the star formation process, since one would expect that any changes in the field geometry (for example, using oppositely directed fields in different regions or a field with a large random component) would tend to \emph{decrease} the importance of magnetic fields in the star formation process, as it would be easier for the fields to reconnect and thus dissipate \citep[e.g.][]{lp96}.

\section{Results}
\label{sec:results}

\subsection{Global cloud evolution}

 The evolution of the global cloud is presented in Figure~\ref{fig:globalcloud}, which shows column density in the cloud at intervals of 0.2 cloud free-fall times (top to bottom) for the five different runs in order of increasing magnetic field strength (left to right). Thus rows correspond to snapshots at a fixed time with varying field strength whilst columns represent a time sequence at a given field strength.
 
  The global cloud evolution at early times ($t/t_{ff} \le 0.4$) is broadly similar in all five cases. Even at $t/t_{ff}=1.2$ (bottom row of Figure~\ref{fig:globalcloud}) the main distinguishing features of the hydrodynamic cloud (overall cloud shape, location of dense regions) remain apparent down to a mass-to-flux ratio of 10. This is so because for mass-to-flux ratios less than 6 (see \S\ref{sec:magfields}, above), the field does not play the dominant role in the gas dynamics of the cloud. However, whilst there are striking differences between the hydrodynamic and strongly magnetised cases, even in the weaker field runs differences due to the magnetic field are apparent.
  
\begin{figure}
\begin{center}
\includegraphics[width=\columnwidth]{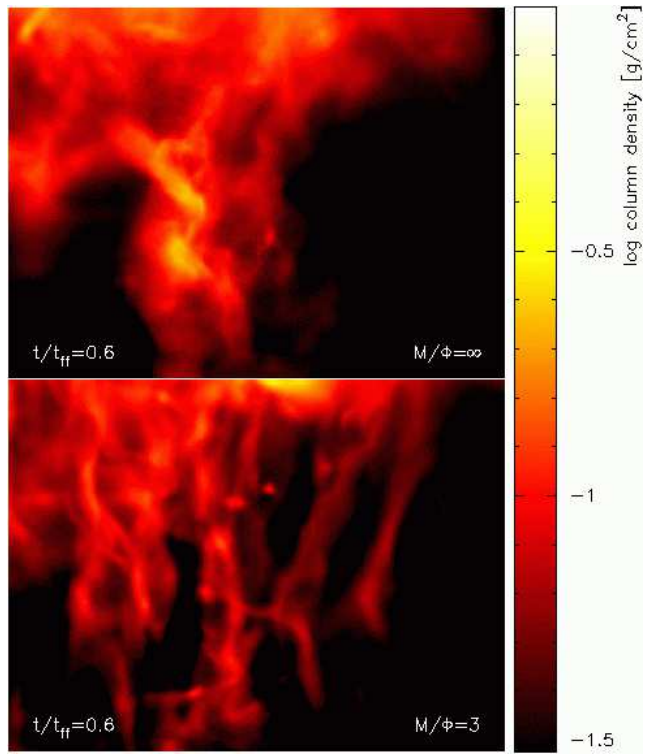}
\caption{Zoomed-in view comparing the outer parts of the cloud in the strong field ($M/\Phi = 3$) run (bottom) to the hydrodynamic run (top) at 0.6 free-fall times. The strong magnetic field run shows filamentary structure in the column density aligned parallel to the field lines (which are approximately vertical -- see Figure~\ref{fig:fieldevol}).}
\label{fig:striations}
\end{center}
\end{figure}

   At early times ($t/t_{ff} < 0.6$, top three rows) there are two main distinguishing characteristics. The first is that the shock structure produced by the initial turbulent velocity field in the dense regions (which appear yellow in the figure) appears smoother and less well-defined than in the hydrodynamic case. We interpret this as being due to the additional pressure support given to the cloud by the magnetic field. A similar effect is observed in hydrodynamic calculations when gas pressure is increased \citep{bb05}. The second notable difference is that the filamentary structure appears \emph{more} filamentary in the less dense regions, particularly evident in the strongest field run ($M/\Phi = 3$) at $t/t_{ff} = 0.4-0.6$ (especially in the lower parts of the cloud in the figure). This increased filamentary structure, or ``stripiness'' is roughly aligned with the large scale magnetic field threading the cloud (see Figure~\ref{fig:fieldevol}). At higher field strengths the field is dominant in these low density regions of the cloud and thus channels the gas flow along the field lines. A close up view of this structure is shown in Figure~\ref{fig:striations}, comparing an enlarged portion of the strongest field run (bottom) at $t/t_{ff} = 0.6$ to the same region in the hydrodynamic run (top). A similar alignment of filamentary structure with magnetic field direction has been recently observed in maps of the Taurus molecular cloud \citep{goldsmithetal05}.

\begin{figure}
\begin{center}
\includegraphics[width=\columnwidth]{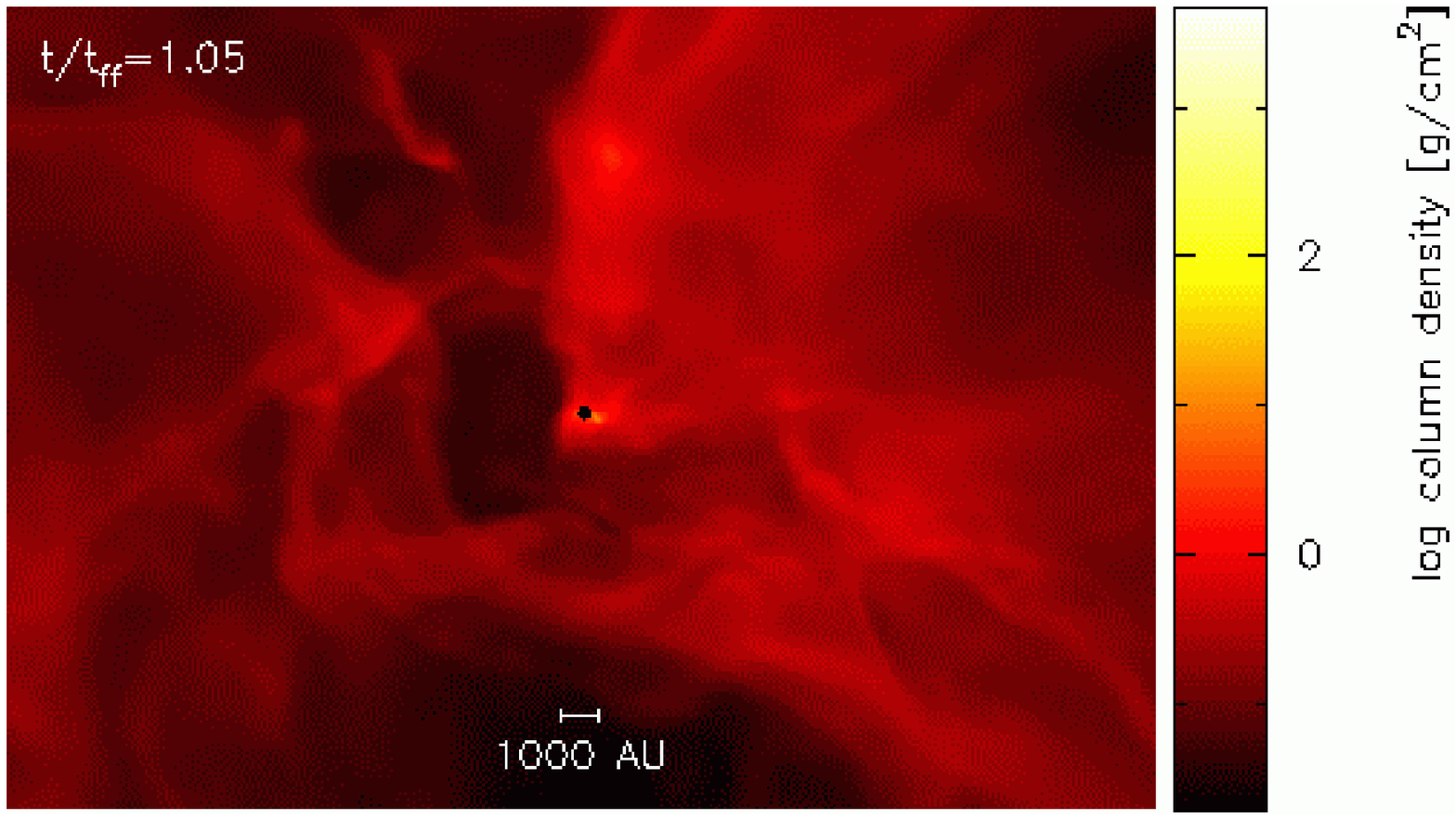}
\includegraphics[width=\columnwidth]{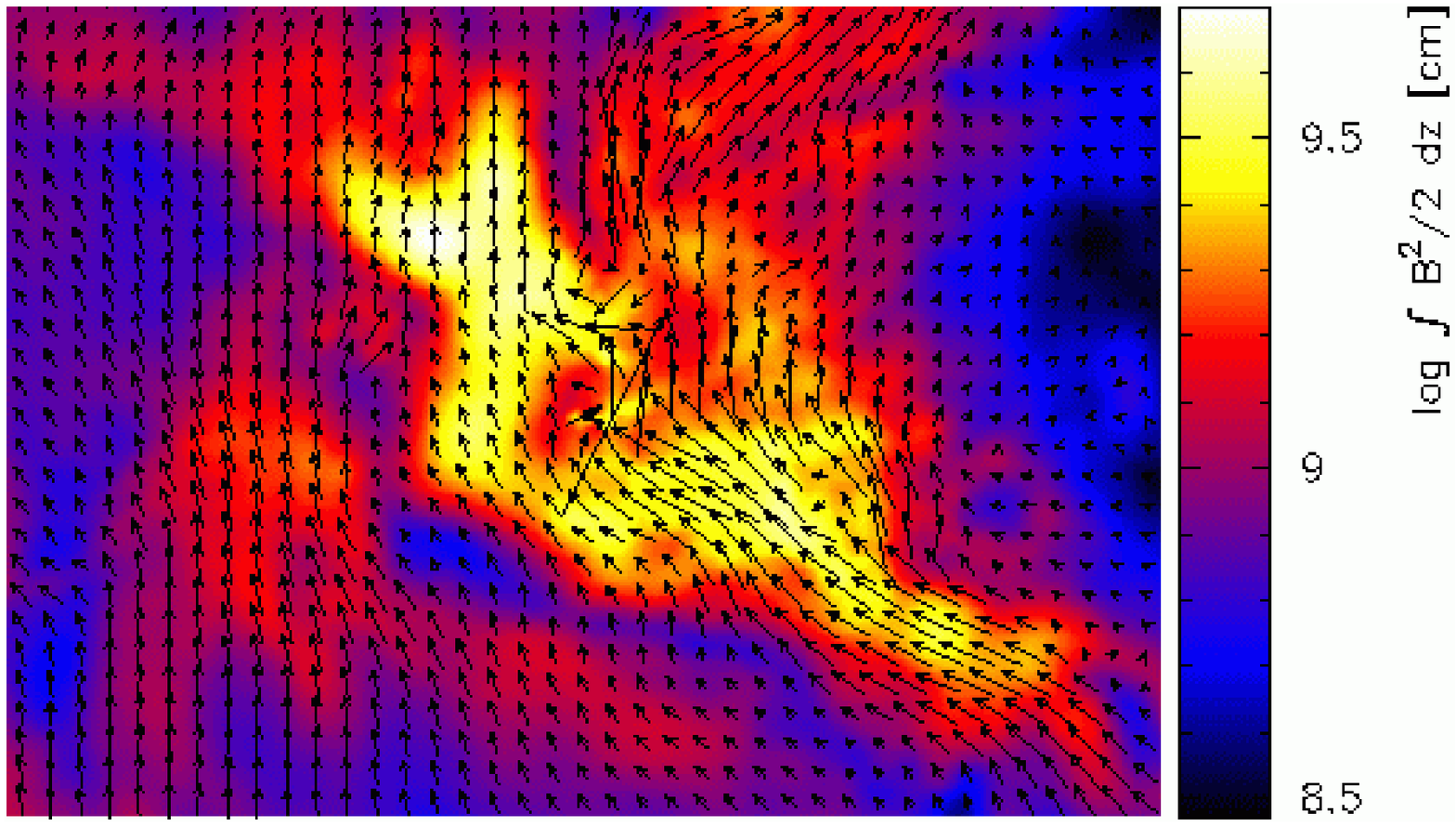}
\caption{Close up view of the void structure in the $M/\Phi=5$ run at $t/t_{ff} = 1.05$, showing column density in the cloud (top) and a rendered plot of the integrated magnetic pressure with overlaid arrows indicating the direction and magnitude of the integrated magnetic field. The lower panel is almost an exact inverse of the upper panel, indicating that the magnetic field is providing the dominant source of pressure in this region.}
\label{fig:magcushion}
\end{center}
\end{figure}

 At later times ($t/t_{ff} > 0.8$) there are further differences in the global cloud evolution. The most obvious of these is that in the $M/\Phi = 5$ and $3$ runs large voids are present in the cloud which are completely absent from the hydrodynamic calculation (e.g. comparing the rightmost panels of the second last and last rows with the hydrodynamic run). These features appear as a result of large scale magnetic flux which remains threaded through the cloud, illustrated further in Figure~\ref{fig:magcushion} which shows a zoomed-in portion of the cloud from the $M/\Phi = 5$ run. The plot shows column density (top panel) together with a plot of the column-integrated magnetic pressure and a map of the integrated magnetic field with strength and direction given by the arrows (bottom panel). The single sink particle which has formed at this point in this simulation is shown in black. Clearly visible is a large void structure to the immediate left of the sink, extending to the upper left and diagonally to the bottom right in the figure. The lower plot, showing the integrated magnetic pressure, appears almost as an inverse of the top panel -- that is, the column density is low where the magnetic pressure is high. Furthermore the magnetic field direction closely traces the void structure visible in the column density plot.
 
  The void is created by material which slides down the magnetic field lines, creating an evacuated region which is unable to be refilled by material perpendicular to the field lines. Since the magnetic flux does not change but the gas pressure decreases, the result is a region where the magnetic pressure is dominant and which prevents material on either side of it from coalescing to form dense structures. The magnetic pressure increase in the lower right part of Figure~\ref{fig:magcushion} is driven further by a ``sandwich'' compression of two dense filaments perpendicular to the field lines (visible above and below the void in the top panel), which squeezes the magnetic field lines and thus increases the magnetic pressure (essentially until the magnetic pressure balances the ram pressure of the filaments), whilst increasing the density proportional only to the one dimensional change in volume. Thus the key to magnetic pressure dominated void creation is a turbulent velocity field which can produce one and two dimensional compressions rather than the isotropic compression produced by gravitational forces.

  The ability of the magnetic field to support parts of the cloud against collapse has significant implications for the star formation rate in the cloud as a whole (and presumably also the overall star formation efficiency) and is discussed further in \S\ref{sec:sfe}.

Finally, a delay in the onset and vigour of star formation is apparent even in this global view at $t/t_{ff} = 1.2$ since stars which have been ejected from their parental envelopes are already visible in the hydrodynamic cloud at this time whilst none are visible in the runs which include a magnetic field. The star formation sequence in each case is discussed further in \S\ref{sec:starform}, below.

\subsection{Magnetic field evolution}
\label{sec:fieldevol}
\begin{figure*}
\begin{center}
\includegraphics[width=\textwidth]{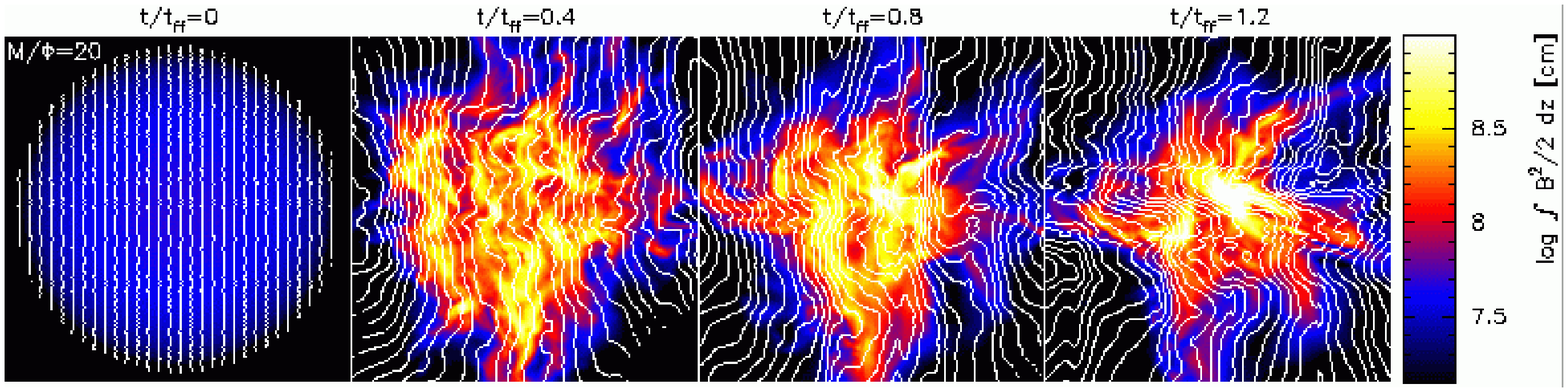}
\includegraphics[width=\textwidth]{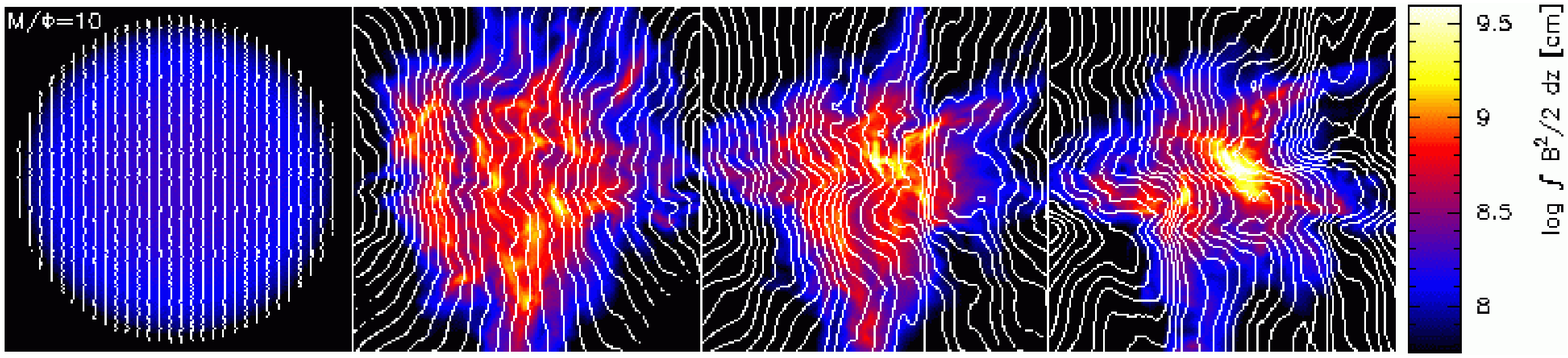}
\includegraphics[width=\textwidth]{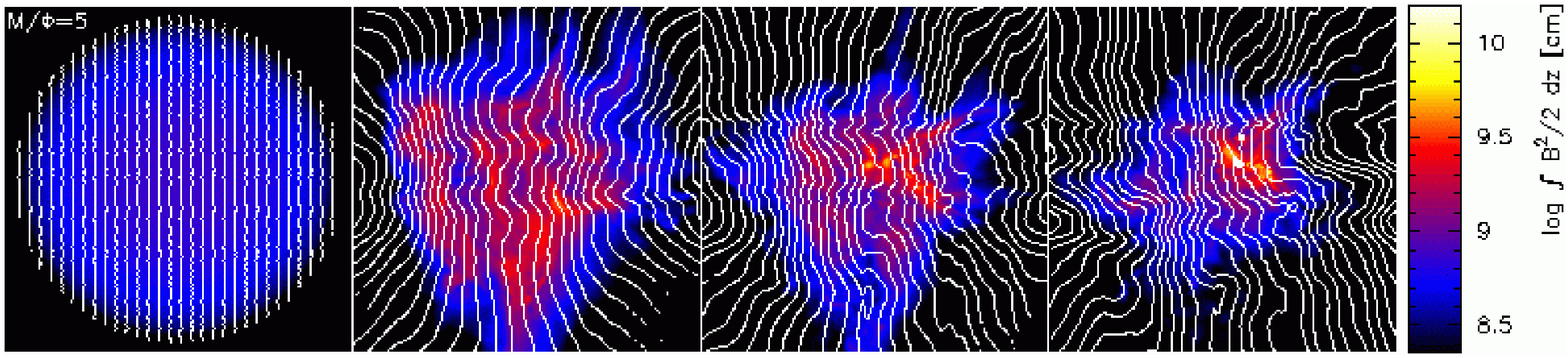}
\includegraphics[width=\textwidth]{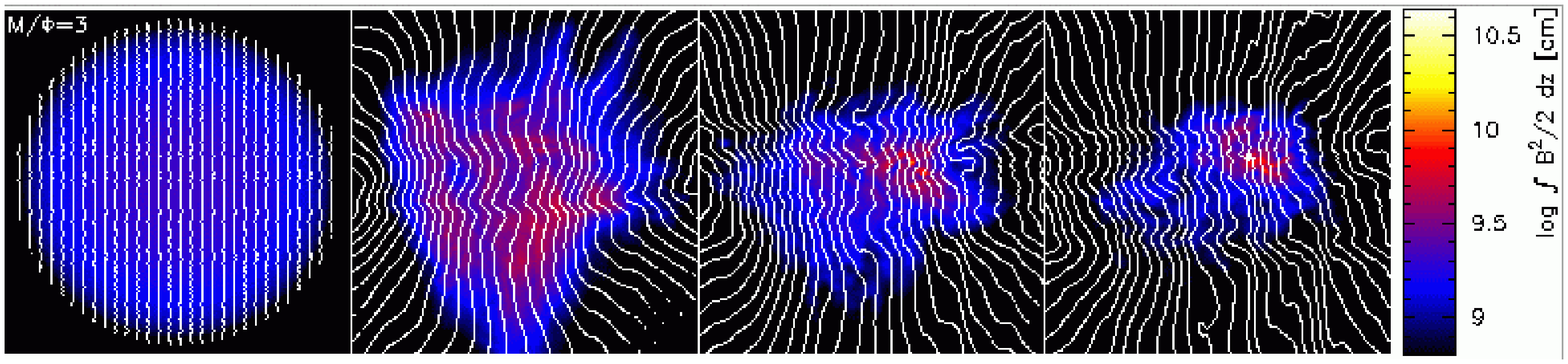}
\caption{Evolution of the magnetic field in each of the magnetised runs ($M/\Phi = 20$, $10$, $5$ and $3$, top to bottom) shown at intervals of 0.4 cloud free-fall times (left to right). Plots show streamlines of the integrated magnetic field direction overlaid on a colour map of the column-integrated magnetic pressure, normalised in each case relative to the initial magnetic pressure (see colour bars). The weaker field runs (top two rows) show strong compression of the magnetic field by the gas, whilst in the stronger field cases (bottom two rows) 
the field is very effective at providing support to the outer regions of the cloud where the column density maps show anisotropic structure parallel to the field lines (Figure~\ref{fig:globalcloud}).}
\label{fig:fieldevol}
\end{center}
\end{figure*}

 The magnetic field in each of the magnetised runs ($M/\Phi = 20, 10, 5$ and $3$) is shown in Figure~\ref{fig:fieldevol} at intervals of $0.4$ cloud free-fall times (left to right). In these plots we show streamlines of the magnetic field direction (column integrated) in the cloud, overlaid on a column-integrated map of the magnetic pressure in the cloud, normalised in each case relative to the initial magnetic pressure. Thus the colour scale illustrates the relative compression of the field in each case. 
 
 In the weaker field runs ($M/\Phi = 20$ and $10$, top two rows) the magnetic field is strongly compressed both by the shocks resulting from the initial turbulent velocity field (most visible at $t/t_{ff} = 0.4$) and subsequently by the gravitational contraction of the cloud ($t/t_{ff} \gtrsim 0.8$, right panels), also resulting in strong distortions of the initially straight magnetic field lines. However even in the weak field cases, whilst the field is significantly distorted by the collapse, the large scale geometry of the field remains imprinted into the cloud by the collapse and the net flux threading the cloud remains apparent even at late times. In fact the large scale structure of the field in the outer regions of the cloud is altered very little as star formation proceeds in the dense central regions.
 
 The relative compression of the field decreases as the field strength increases (ie. comparing snapshots within the same column) and in the stronger field runs ($M/\Phi = 5$ and $3$, bottom two rows) the field geometry remains largely uniform as the collapse proceeds, with only a relatively small compression of the magnetic field. In these cases the magnetic field is able to impart significant directionality to the gas motions -- particularly in the outer parts of the cloud, by channelling material along magnetic field lines (the effects of which are clearly visible in the column density plots for these runs shown in Figure~\ref{fig:globalcloud}). The anisotropy of turbulent motions in the presence of a magnetic field is a clear prediction of MHD turbulence theory (e.g. \citealt{gs95}). This also leads to the tantalising possibility that it may be possible to infer and thus map both magnetic field strength and direction in molecular clouds by measuring anisotropy in interferometric velocity maps \citep{vos03}.

  Finally, Figure~\ref{fig:fieldevol} also illustrates how the boundary condition on the magnetic field is treated in the calculations (discussed above in \S\ref{sec:magfields}). Initially (left panels) the field is defined only on the particles but remains uniform at the boundary because the gradient in the Euler potentials is computed exactly to linear order regardless of the particle distribution. By $t/t_{ff} = 0.4$ (second panel), the outer layers of the cloud have expanded and thus provide an external medium into which the magnetic field remains anchored at later times.

\subsection{Star formation sequence}
\label{sec:starform}

\begin{figure*}
\begin{center}
\includegraphics[width=\textwidth]{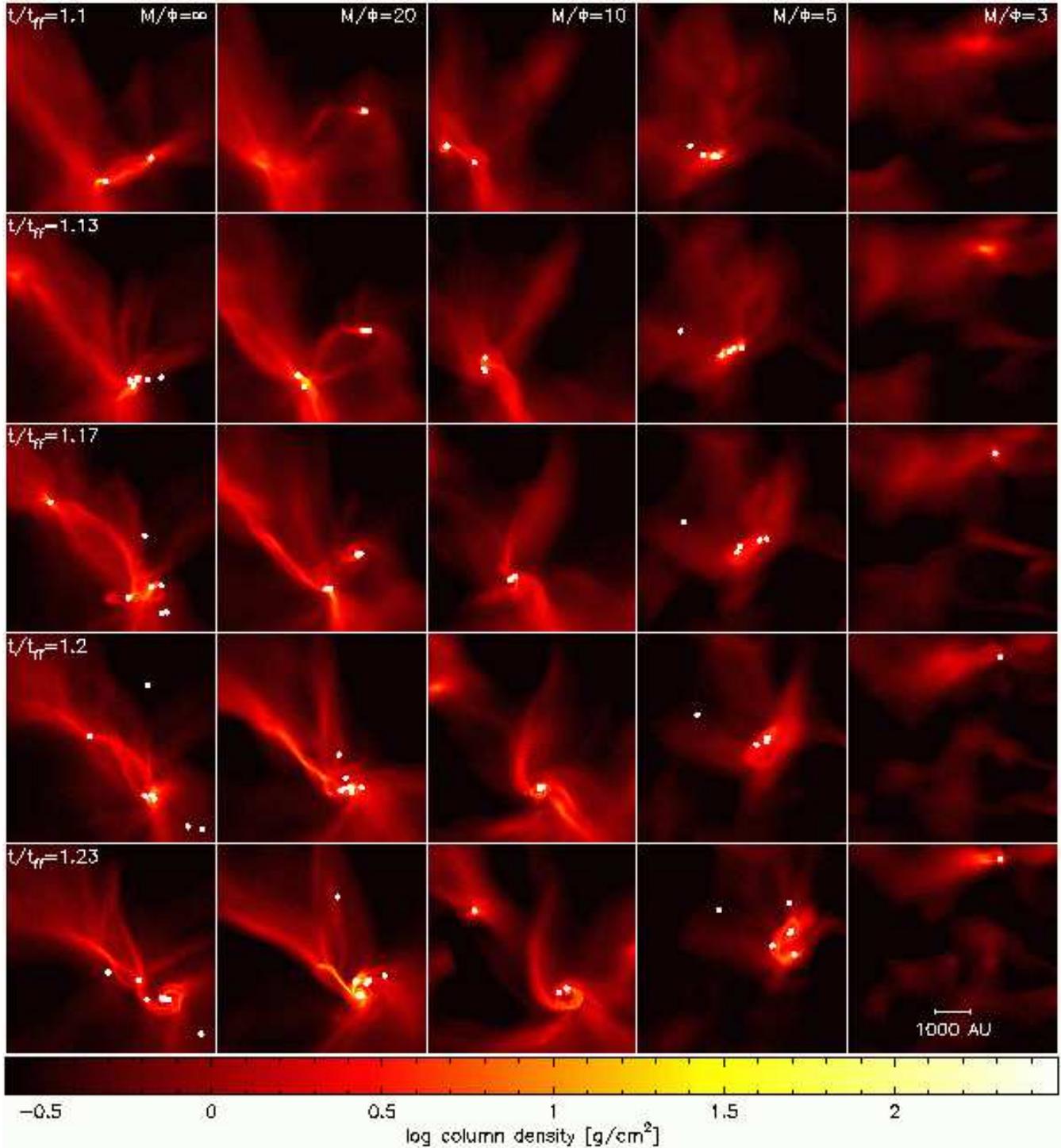}
\caption{Close up view of the star formation sequence in each of the 5 runs, shown at intervals of 0.032 cloud free-fall times (top to bottom, continued in Figure~\ref{fig:starform2}) and in order of increasing magnetic field strength (left to right). Whilst the hydrodynamic run collapses in three main regions which subsequently merge, the magnetised runs show delayed collapse in some or all of these regions, leading to more quiescent dynamics and an almost complete suppression of star formation in the $M/\Phi=3$ case.}
\label{fig:starform1}
\end{center}
\end{figure*}

\begin{figure*}
\begin{center}
\includegraphics[width=\textwidth]{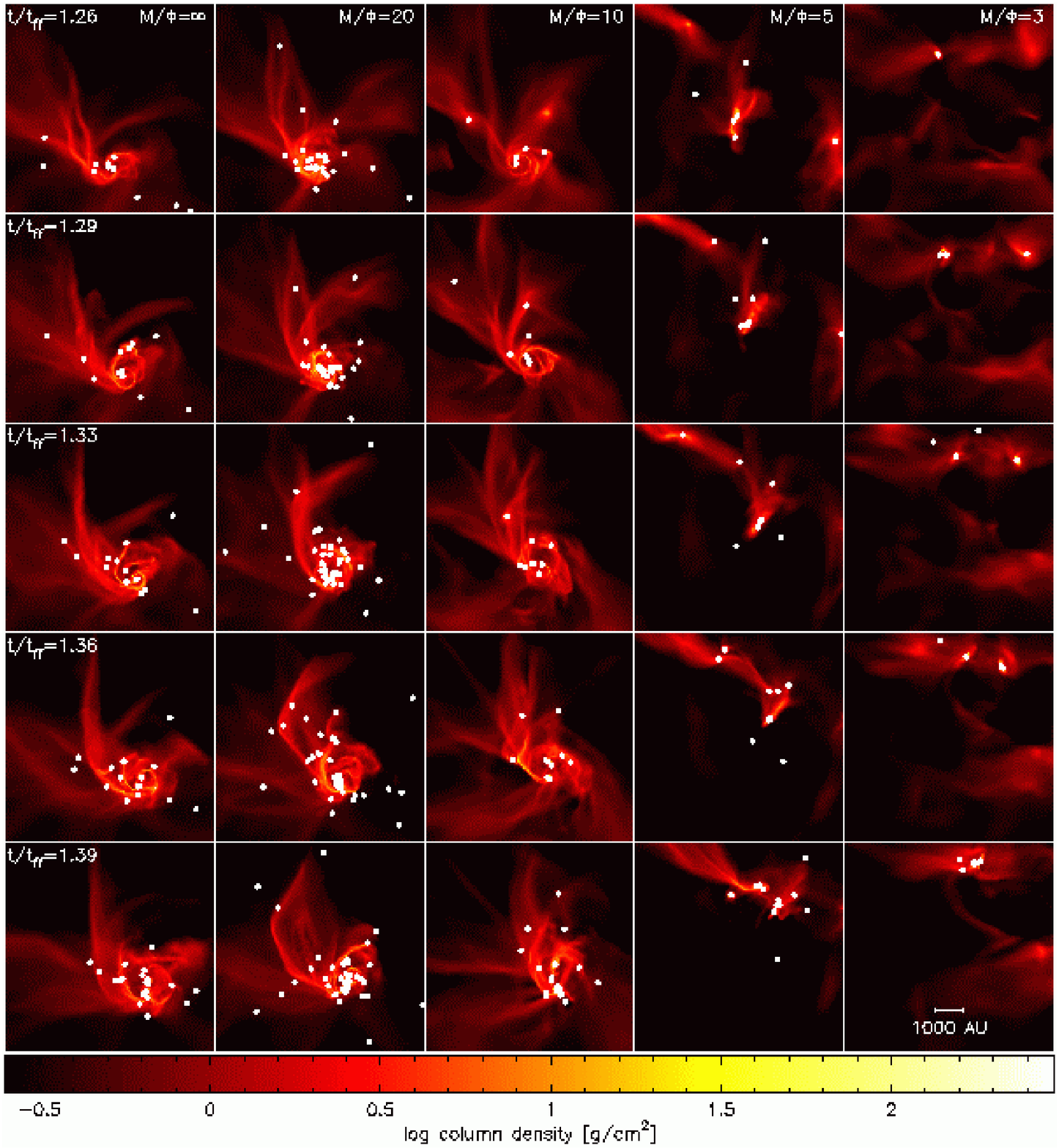}
\caption{As in Figure~\ref{fig:starform1} but showing the later time evolution ($t/t_{ff} \ge 1.26$) in a slightly wider view (dimensions $7219 \times 7219$AU). The hydrodynamic and $M/\Phi = 20$ runs both contain massive disc fragmentation which is delayed in the $M/\Phi=10$ run and completely absent from the $M/\Phi = 5$ and $M/\Phi=3$ runs which show a much more subdued star formation sequence and fewer violent ejections.}
\label{fig:starform2}
\end{center}
\end{figure*}

 The star formation sequence in each of the five runs is presented in Figures~\ref{fig:starform1} and \ref{fig:starform2} although it is best appreciated by viewing animations of each simulation\footnote{http://www.astro.ex.ac.uk/people/dprice/research/mcluster/}. The figures show snapshots on a zoomed-in portion of the cloud (dimensions $5156 \times 5156$AU and $7219 \times 7219$AU in Figures~\ref{fig:starform1} and \ref{fig:starform2} respectively) at intervals of 0.032 free-fall times throughout the evolution. We again caution that the results in the MHD cases should be taken as an upper limit on the degree of star formation expected in ideal MHD due to the high numerical resistivity present on small scales (although this may not be completely unrepresentative since we have also neglected physical diffusion processes such as ambipolar diffusion). Nonetheless, the figures serve to starkly illustrate how the effect of the magnetic field on large scales can have a significant influence on both the degree and manner of star formation which occurs in the cloud.

\subsubsection{Hydrodynamic run}
  In the hydrodynamic case (leftmost column), star formation initiates in three dense cores (two of which are shown in the $t/t_{ff}=1.1$ panel, the other collapses to the top left of this figure and is visible at $t/t_{ff}=1.17$). The two protostars shown in the $t/t_{ff}=1.1$ panel accrete gas rapidly to reach masses of around $0.1$ and $0.3 M_{\odot}$. The accretion flow forms massive discs around each of these stars, both of which subsequently fragment to give a triple and quadruple (double binary) system, respectively, from each of which low mass members are ejected (though some ``dance'' around the combined potential of the two systems before being subsequently ejected). The two systems (initially separated by $\sim 2000$ AU) fall towards each other on elliptical orbits and eject several protostars at each of two periastron passages (seen at closest approach in the panel shown at $t/t_{ff} = 1.13$) before the two systems enter a more circular orbit around each other (see panel at $t/t_{ff} = 1.2$). The dynamics is then dominated by this ``binary'' system (that is, with two main concentrations of mass orbiting each other) surrounded by a circum``binary'' accretion flow. One of the stars in one half of the binary system briefly forms a circumstellar disc ($R\sim 125 AU$) before it is destroyed by dynamical interactions with low mass stars in the process of being ejected from the system.
  
   The system receives a strong perturbation when the third of the first three dense cores (visible in the upper left of the $t/t_{ff}=1.2$ panel falling towards the binary system), together with the disc which forms around it during the infall, crashes through the main system rather like a cannonball and ejects two protostars in a spectacular ``billiard-ball'' style encounter. Thus, at $t/t_{ff}=1.26$ (first panel in Figure~\ref{fig:starform2}) the dynamics transitions from mostly a two-body system (plus perturbations from multiple lower mass members) to a ``triple'' system (ie. three interacting dense cores) surrounded by a large and massive circum-triple disc which fragments into multiple single and binary systems (subsequent panels in Figure~\ref{fig:starform2}, $t/t_{ff} > 1.33$).

\subsubsection{$M/\Phi = 20$}
\label{sec:mf20sf}
 At early times ($t/t_{ff} \lesssim 0.6$) the highly supercritical run ($M/\Phi = 20$) evolves almost identically to the hydrodynamic case (Figure~\ref{fig:globalcloud}, second column), though the filamentary structure appears marginally smoother because of the increased pressure provided by the (albeit weak) magnetic field. The slight additional pressure provided by the magnetic field also changes the initial star formation sequence, as two of the three dense cores form slightly later than in the hydrodynamic case. More importantly the third core does not collapse ($t/t_{ff} = 1.1 - 1.17$ in Figure~\ref{fig:starform1}, second column) until just before it has merged into the main concentration of mass. This slight change in the star formation sequence has a dramatic effect on the results, since instead of separate multiple systems forming (as in the hydrodynamic case), in this case the accretion streams coalesce into one very massive disc which subsequently fragments (Figure~\ref{fig:starform2}, $t/t_{ff}=1.26$ onwards) and undergoes rapid and vigourous star formation ($t/t_{ff}=1.26-1.39$ in Figure~\ref{fig:starform2}). Though we caution that such disc fragmentation may be an artifact of the barytropic equation of state employed in the calculations (see discussion in \S\ref{sec:eos}), the difference between this run and the hydrodynamic case serves to illustrate the chaotic nature of turbulent star formation, in that even the introduction of a weak magnetic field can produce a dramatic difference in the results. 

\begin{figure*}
\begin{center}
\includegraphics[width=0.45\textwidth]{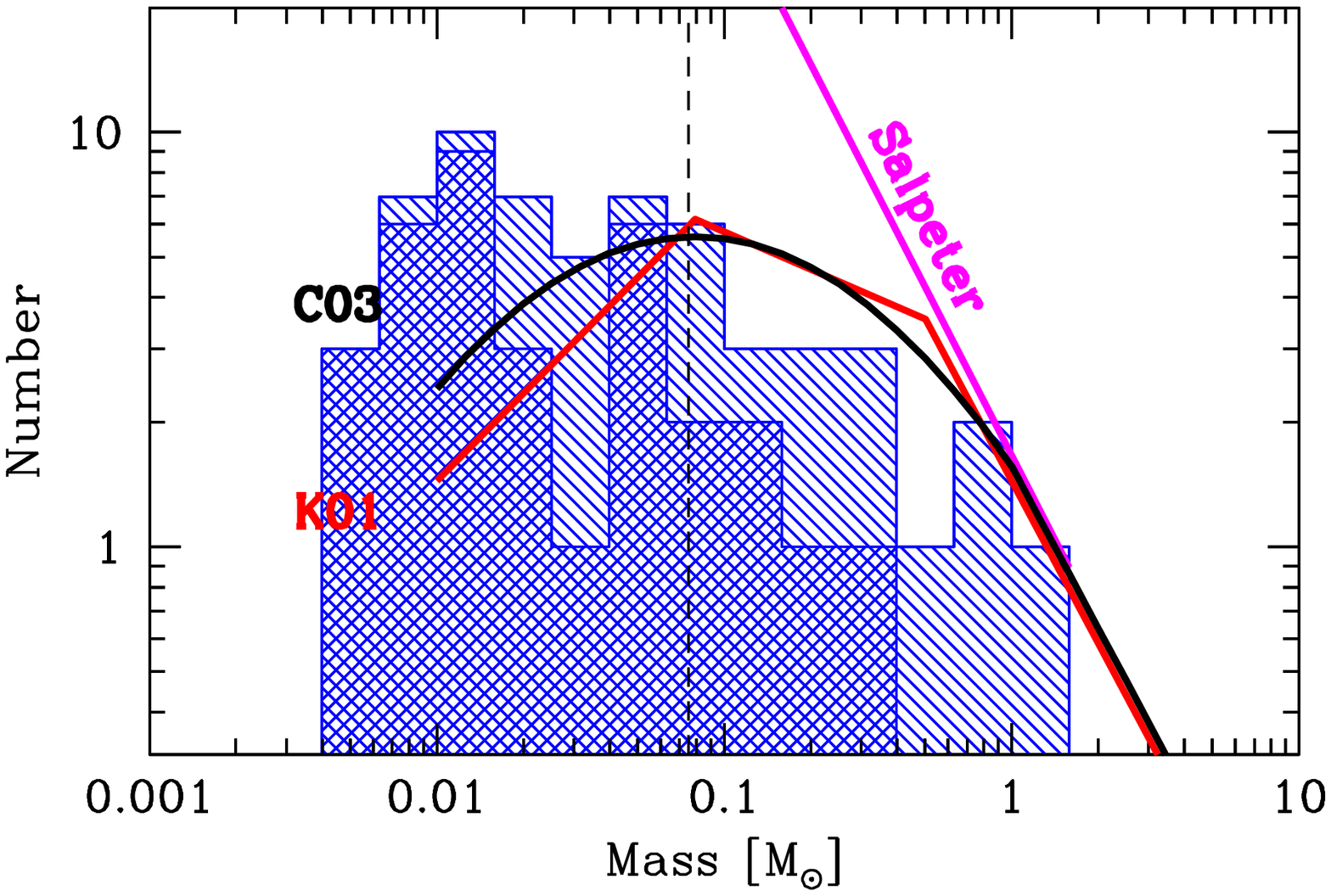}
\includegraphics[width=0.45\textwidth]{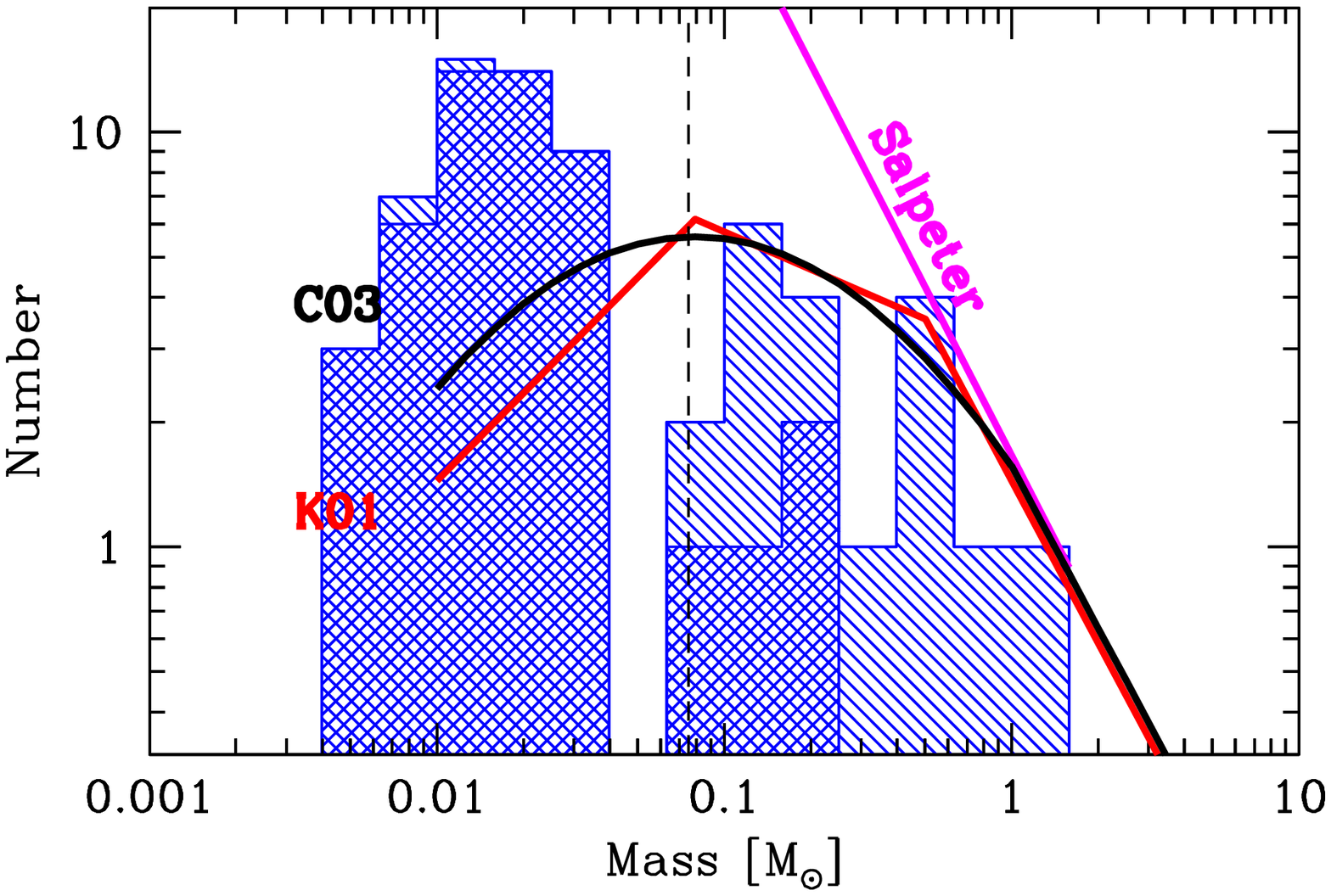}
\includegraphics[width=0.45\textwidth]{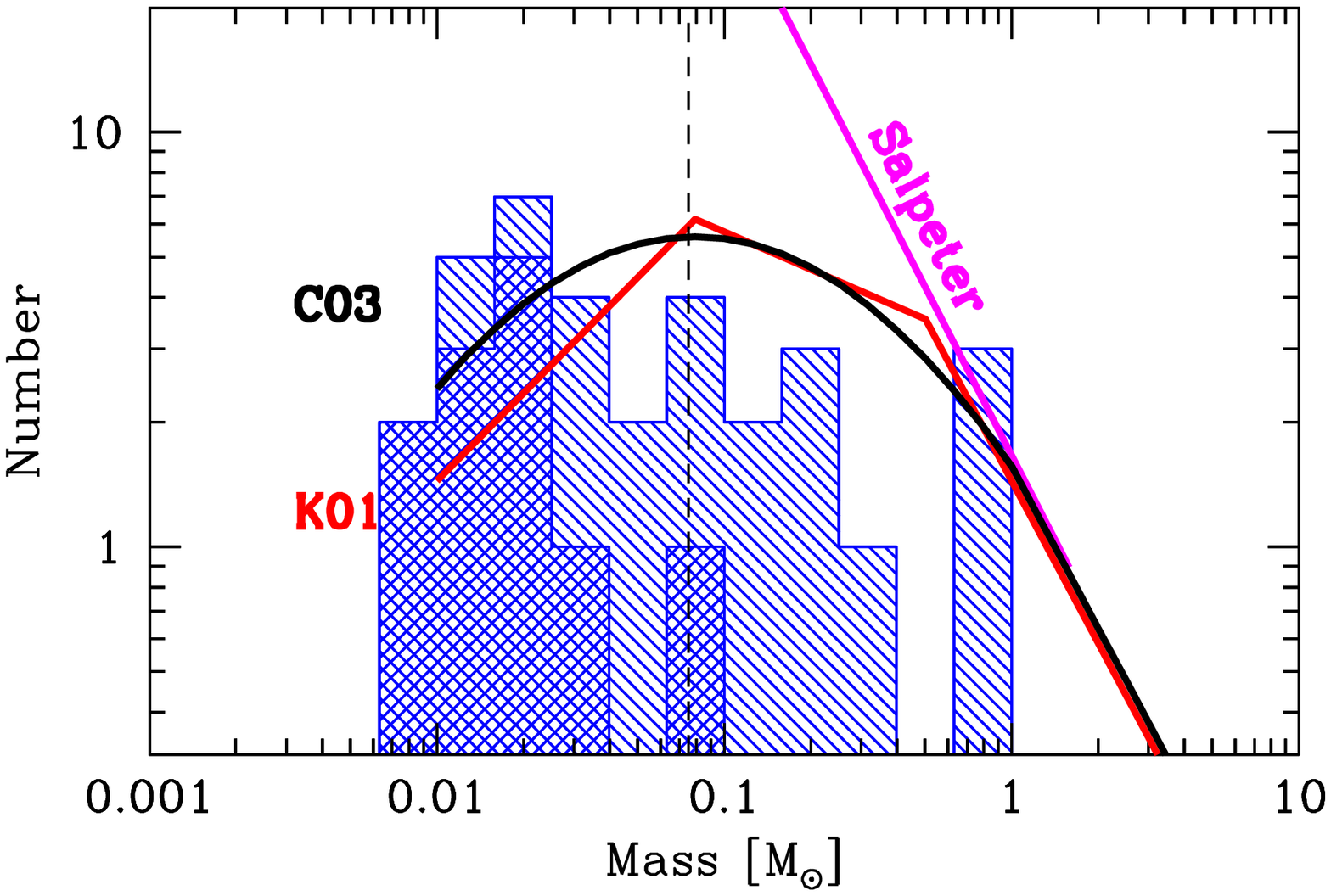}
\includegraphics[width=0.45\textwidth]{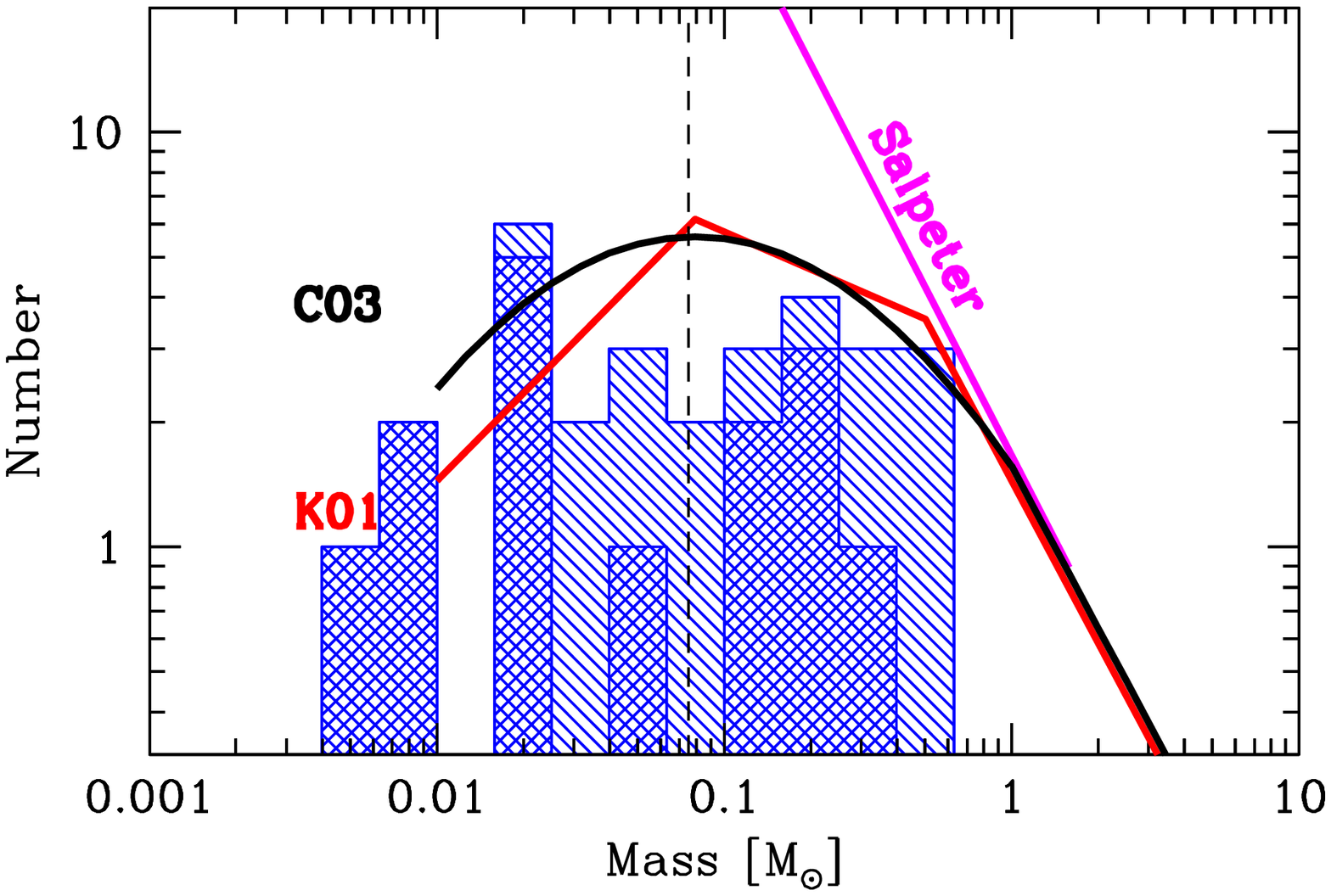}
\includegraphics[width=0.45\textwidth]{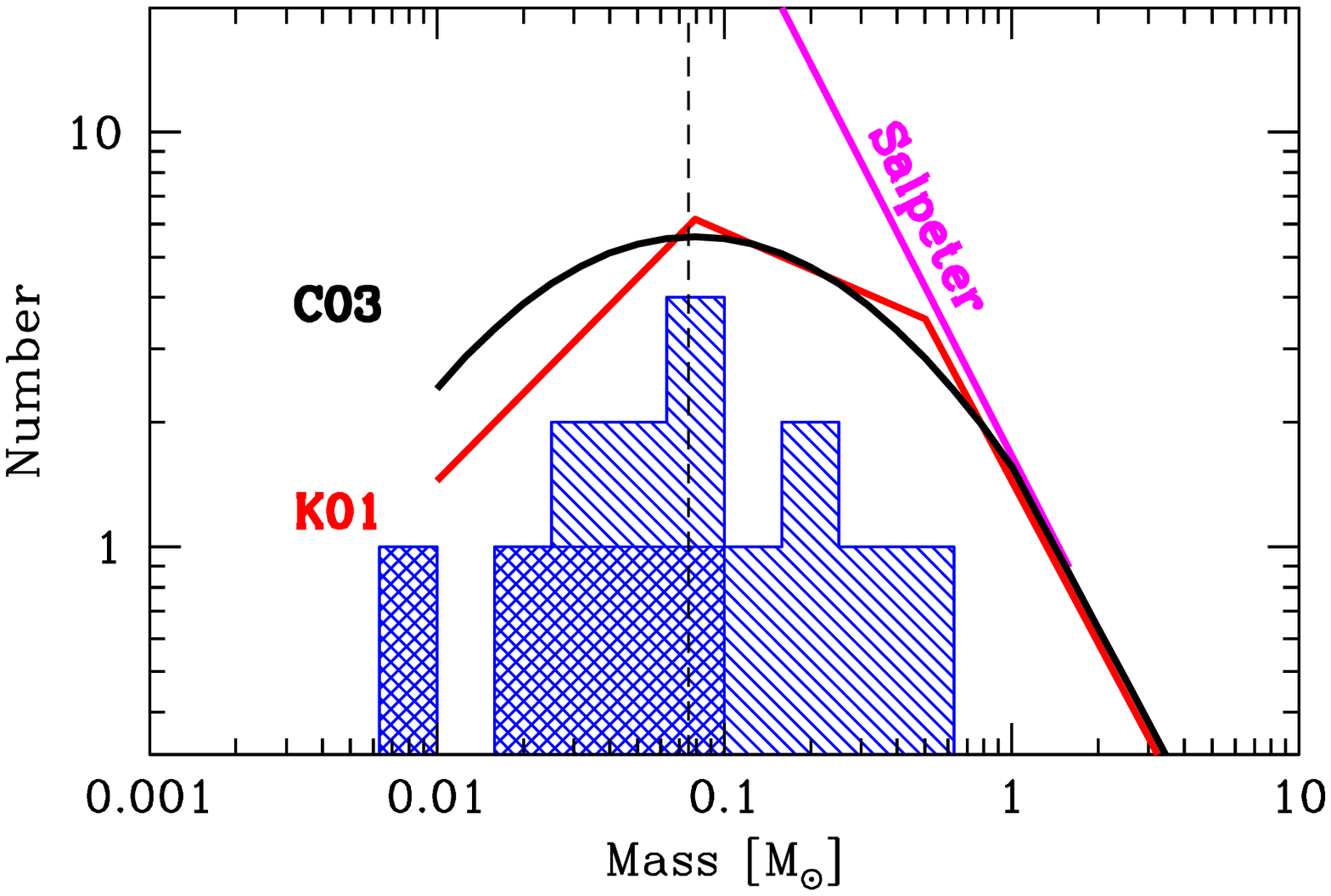}
\includegraphics[width=0.45\textwidth]{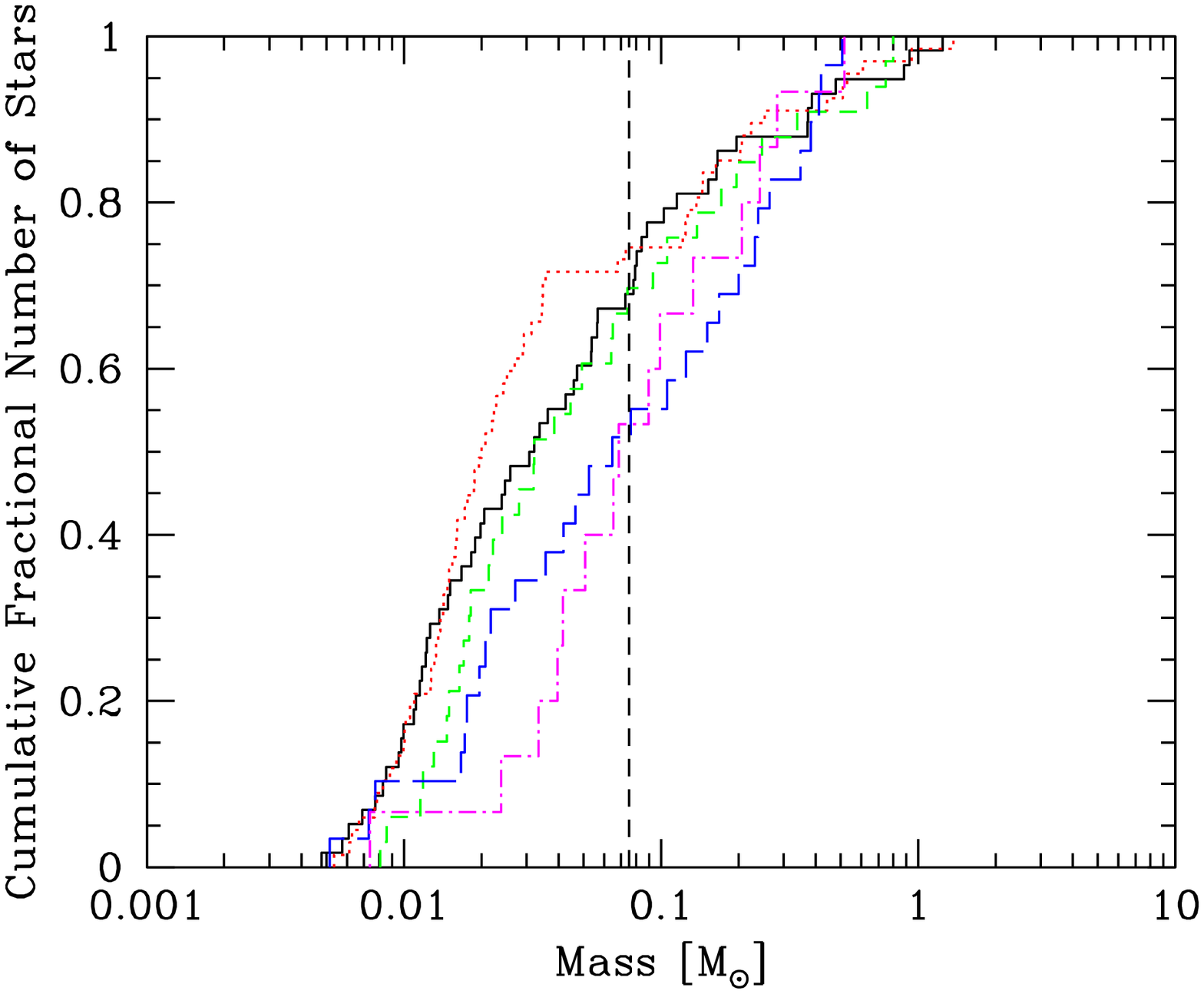}
\caption{Initial mass functions at 1.5 free fall times for each of the five runs, in order of increasing magnetic field strength (left to right, top to bottom) and the cumulative fractional number of stars as a function of mass in all four cases (bottom right panel), with lines corresponding to the hydrodynamic run (black, solid), $M/\Phi=20$ (red, dotted), $M/\Phi=10$ (green, dashed), $M/\Phi=5$ (blue, long-dashed) and $M/\Phi=3$ (magenta, dot-dashed). The vertical dashed line in each case indicates the characteristic mass in the hydrodynamic run.}
\label{fig:imfs}
\end{center}
\end{figure*}

\subsubsection{$M/\Phi = 10$}
 The star formation sequence in the $M/\Phi = 10$ run differs further from the hydrodynamic case. At $t/t_{ff}=1.17$ in this run (central panel of Figure~\ref{fig:starform1}) only the densest of the three initial cores in the hydrodynamic run has collapsed, though the core visible to the upper left of the main star formation at $t/t_{ff}=1.17$ in the hydrodynamic case can be seen to form, though later ($t/t_{ff}=1.23$, centre panel of last row in Figure~\ref{fig:starform1}) and at a much greater distance from the densest region. The core which forms to the upper right in the hydrodynamic case is completely suppressed in the $M/\Phi = 10$ run until around $t/t_{ff}=1.29$ (visible forming in the centre panel of the top two rows in Figure~\ref{fig:starform2}). As a result the subsequent disc in the core of the cluster (which fragments spectacularly in the $M/\Phi = 20$ run) is at earlier times much less massive and forms only an accreting binary system ($t/t_{ff} = 1.29$ in Figure~\ref{fig:starform2}), though the mass accretion onto this system at later times ($t/t_{ff} > 1.33$, centre panel of last three rows in Figure~\ref{fig:starform2}) causes further fragmentation.

\subsubsection{$M/\Phi = 5$}
 The star formation sequence in the $M/\Phi = 5$ run (Figure~\ref{fig:starform1}, fourth column) is almost unrecognisable compared to the hydrodynamic case. The first fragmentation occurs in this case in a disc which appears edge-on in Figure~\ref{fig:starform1} ($t/t_{ff} = 1.1$ panel) -- that is, perpendicular to the global magnetic field direction. Whilst this disc fragments to form a multiple system from which a brown dwarf is ejected, the subsequent accretion and thus star formation occurs at a dramatically reduced level in this run. Thus whilst inside the dense core there are still violent interactions between stars, the core itself is generally starved of new material by the lower accretion rate and is unable to form any more than a handful of stars. Also, without the dramatic encounters with protostars formed at greater distances, the ejection of low mass protostars is much less efficient. An increase in star formation activity occurs at $t/t_{ff} > 1.33$ as the severely delayed collapse of the upper left core occurs (Figure~\ref{fig:starform2}). The role of the magnetic field in suppressing accretion from the cloud is quantified and discussed further below. 

\subsubsection{$M/\Phi = 3$}
 In the $M/\Phi = 3$ run the collapse is strongly channelled along the magnetic field lines (Figures~\ref{fig:globalcloud} and \ref{fig:fieldevol}) and there is relatively little compression of the global magnetic field by the gas (Figure~\ref{fig:fieldevol}). Star formation in the cloud (Figure~\ref{fig:starform1}, rightmost column) is strongly suppressed -- only a single, isolated core has collapsed by $t/t_{ff}=1.23$ (last panel in Figure~\ref{fig:starform1}). It is not until $t/t_{ff}=1.29$ (Figure~\ref{fig:starform2}) that a second core collapses and even then the two remain spatially isolated until around $t/t_{ff}=1.39$ (last panel of Figure~\ref{fig:starform2}). The manner in which star formation proceeds in the cloud is completely different from both the hydrodynamic, weak field, and even the $M/\Phi = 5$ case. Features are visible in the column density which are clearly not gravitational in origin (for example the ``streamer'' visible at $t/t_{ff} > 1.29$ in Figure~\ref{fig:starform2}).

\subsection{Initial mass functions}
 The initial mass function (IMF) evaluated in each of the runs is shown in Figure~\ref{fig:imfs}, in order of increasing magnetic field strength (left to right, top to bottom), in each case evaluated at 1.5 cloud free-fall times. The dark hatched portion of the histogram indicates the stars that have ceased accreting from the global cloud whilst the light hatched portion indicates those stars that are still accreting at $t/t_{ff}=1.5$. For comparison the slope of the \citet{salpeter55} IMF is plotted (purple line) as well as the IMFs covering the substellar population determined by \citet{chabrier03} and \citet{kroupa01}.

Whilst the low number of stars and brown dwarfs formed overall in the simulations precludes a detailed evaluation of the effect of magnetic fields on the IMF, there is a hint of some general trends. The first is in the overall normalisation -- that is, comparing the total number of objects formed in each case. For example, comparing the strongest field run ($M/\Phi=3$) in Figure~\ref{fig:imfs} (left panel, bottom row) to the hydrodynamic and weak field ($M/\Phi=20$) cases (top row) it is clear that fewer objects are formed overall in the strong field case.
 
 A further question, and one which can only be tentatively approached given the limitations of the present simulations, is whether or not magnetic fields have an influence on the shape of the IMF. The lower right panel in Figure~\ref{fig:imfs} shows the cumulative fractional number of stars as a function of mass for each of the 5 runs of varying field strength. Whilst we caution that the statistics are low, the general direction is a trend towards fewer low mass objects with increasing field strength. This is better illustrated by looking at cruder statistics such as the ratio of stars to brown dwarfs formed in each of the simulations, given in Table~\ref{tab:starbdratio}. From the table we see that in the hydrodynamic and weak field runs there tends to be an excess in the number of brown dwarfs relative to the number of stars, by as much as a factor of 3. In contrast, in the magnetically dominated runs, we find roughly equal numbers of stars and brown dwarfs. We attribute this to the suppression of accretion in the stronger field runs, leading to fewer protostars, fewer dynamical interactions and thus fewer ejections of low mass objects.
\begin{table}
\begin{center}
\begin{tabular}{|l|l|l|l||}
\hline
$M/\Phi$ & $N_{BDs}$ & $N_{stars}$ & ratio \\
\hline
$\infty$ & 44 & 14 & 3.14 \\
$20$ & 51 & 18 & 2.83 \\
$10$ & 22 & 11 & 2.0 \\
$5$ & 15 & 14 & 1.07 \\
$3$ & 8 & 7 & 1.14 \\
\hline
\end{tabular}
\caption{Ratio of brown dwarfs to stars formed in each of the runs. Whilst there are at present only low number statistics for both populations, there is some indication of a trend towards fewer brown dwarfs relative to stars in the presence of strong magnetic fields.}
\label{tab:starbdratio}
\end{center}
\end{table}

\subsection{Star formation rate}
\label{sec:sfe}
\begin{figure}
\begin{center}
\includegraphics[angle=270,width=\columnwidth]{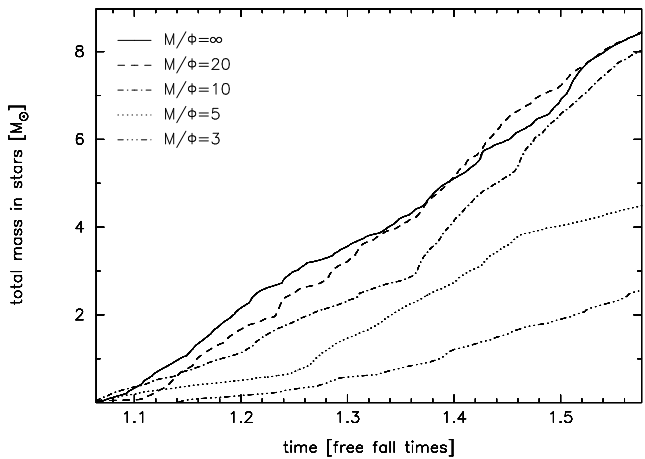}
\caption{Effect of the magnetic field on the star formation rate in each of the 5 runs. The plot shows the total mass accreted onto sink particles as a function of time in each of the calculations. A clear trend is visible in which the accretion from the cloud is increasingly suppressed as the magnetic field strength increases. The strongest field run ($M/\Phi=3$) shows a $75\%$ reduction in the total mass accreted onto stars at $t/t_{ff} = 1.5$ compared to the hydrodynamic case.}
\label{fig:massinstars}
\end{center}
\end{figure}

 The effect of the magnetic field in suppressing accretion from the global cloud onto the star forming cores is quantified in Figure~\ref{fig:massinstars}, which shows the total mass in stars (that is, the total mass of all gas accreted onto sink particles) as a function of time in units of free-fall times for the 5 runs which form stars (ie. up to $M/\Phi = 3$, with runs indicated by the legend). It is clearly apparent from this figure that the mass accretion rate strongly anti-correlates with magnetic field strength. Even with a very  weak magnetic field ($M/\Phi = 20$), the accretion rate is clearly lower than the hydrodynamic case up until the disc fragmentation which occurs at around $t/t_{ff}= 1.3$ (Figure~\ref{fig:starform1}).
 
  In the $M/\Phi = 10$ case the accretion rate at early times (up to $\sim t/t_{ff} = 1.35$) is around half of that in the hydrodynamic run -- $\dot{M} \sim 8.5 M_{\odot}/t_{ff}$ compared to $\dot{M}_{Hyd} \sim 16 M_{\odot}/t_{ff}$. The strong-field runs ($M/\Phi=5$ and $M/\Phi=3$) both show very low initial accretion rates $\dot{M} \sim 3-4 M_{\odot}/t_{ff}$. The difference between the two is that the accretion rate in the $M/\Phi = 5$ run increases dramatically at around $t/t_{ff}=1.25$ as two relatively distant regions of the cloud undergo gravitational collapse (further out than the regions shown in Figure~\ref{fig:starform1}), whereas this does not occur in the $M/\Phi = 3$ run. In fact the accretion rate between $t/t_{ff} = 1.25-1.46$ in the $M/\Phi=5$ run, $\dot{M} \sim 14 M_{\odot}/t_{ff}$, is only slightly lower than the average hydrodynamic rate. However the low initial and later accretion rates mean that by $t/t_{ff} = 1.5$ there is around half of the mass in stars in the $M/\Phi = 5$ run compared to the hydrodynamic case ($\sim 4 M_{\odot}$ compared to $\sim 8 M_{\odot}$, or a 50\% reduction). In the very strong field run the effect is even more dramatic -- by $t/t_{ff} = 1.5$ there is only around one quarter of the mass in stars compared to the hydrodynamic case ($\sim 2 M_{\odot}$ compared to $\sim 8 M_{\odot}$, or a 75\% reduction in the mass converted to stars).

\begin{figure}
\begin{center}
\includegraphics[angle=270,width=\columnwidth]{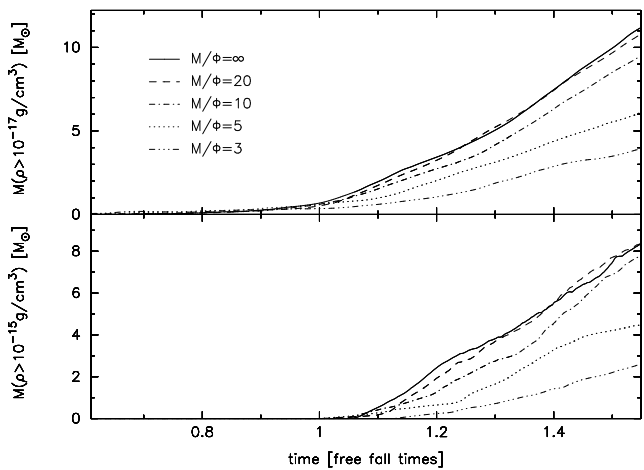}
\caption{Mass above a particular density in the cloud as a function of free-fall time for the 5 runs (lines as indicated in the legend), showing the results for density values of $10^{-17}$g/cm$^{3}$ (top panel) and $10^{-15}$g/cm$^{3}$ (bottom panel). Again a clear trend is apparent towards lower mass infall rates with increasing magnetic field strength.}
\label{fig:massaboverho}
\end{center}
\end{figure}

 As an illustration of the effect of magnetic fields in preventing lower density gas from collapsing (and thus the effect on larger scales) as suggested by \citet{kt07} we plot the mass above a particular density in the cloud as a function of time in Figure~\ref{fig:massaboverho}. The plot shown $M(>\rho)$ for density values of $\rho = 10^{-17}$g/cm$^{3}$ (top panel) and $\rho = 10^{-15}$g/cm$^{3}$ (bottom panel). A similar trend towards lower mass infall rates with increasing magnetic field strength is also visible in these plots.

\section{Discussion}
\label{sec:discussion}
 We have performed a study of how magnetic fields affect the large scale collapse of turbulent molecular clouds to form star clusters, computing a range of models with mass-to-flux ratios ranging from highly to moderately supercritical (with a corresponding range in the ratio of gas to magnetic pressure, $\beta$). Whilst even the weakest field runs show differences when compared to the hydrodynamic case (e.g. lower accretion rates, different star formation sequences), strong differences in the gas dynamics were found to be present in the runs where the magnetic pressure dominates the gas pressure (ie. $\beta < 1$), in the form of magnetically-supported voids, column density striations due to anisotropic turbulent motions and much lower accretion rates from the global cloud. The implication of these results for both observations of star-forming molecular clouds and for our theoretical understanding of star cluster formation are discussed below.

\subsection{Relevance to observations}
 Zeeman measurements of magnetic fields in molecular clouds \citep{crutcher99,bourkeetal01} suggest typical field strengths of $B\sim 10\mu$G (and generally $B \lesssim 30\mu$G) for regions with $T \sim 10K$, mass-to-flux ratios which are supercritical by a factor of $\sim 2-3$, marginally (or perhaps firmly, see \citealt{padoanetal04}) super-Alfv\'enic turbulent velocities and $\beta \sim 0.03-0.6$ (similar values for $\beta$ and the ratio of turbulent to Alfv\'enic velocities are also inferred by \citealt{ht05} in the wider Cold Neutral Medium). The inference therefore is that the most realistic of our calculations are actually the two strongest field runs ($M/\Phi = 3$ and $M/\Phi = 5$). Given that this is the case, we should expect that \emph{all} of the magnetically-driven features observed in these simulations to also be present in observed molecular clouds.
 
  Recent first results from a large-scale mapping project of the Taurus molecular cloud complex by \citet{goldsmithetal05} report ``ring, arc and bubble-like features'' and ``striated structure... correlated with the magnetic field direction'' in $^{12}$CO maps. In fact almost the whole of the low density material in Taurus in the \citet{goldsmithetal05} map appears striated parallel to the global magnetic field threading the cloud as mapped from polarization measurements, as was found to occur in the simulations in the low-density outer regions of the clouds. This is suggestive of a low plasma $\beta$ in these regions, since we only find striated structure in the two strongest field simulations with $\beta < 1$ (and most prominently in the $M/\Phi = 3$ case where $\beta = 0.3$). This is in broad agreement with the polarisation measurements of \citet{crutcher99} giving lower limits of $\beta \gtrsim 0.06$ for regions within Taurus.

  One of the most striking features of the magnetised collapse simulations is the appearance of large-scale magnetic-pressure supported voids produced by the large scale magnetic flux tubes threading the cloud. In these regions the integrated magnetic pressure appears to anti-correlate with the column density of the cloud (Figure~\ref{fig:magcushion}). Turning again to the observations of Taurus, \citet{goldsmithetal05} report a ``very interesting feature'' at $4^{h}30^{m} + 25^{\deg}$ which ``appears as a hole'', where it appears that ``some agent has been responsible for dispersing the molecular gas''. We propose that this may be a magnetic-pressure driven feature. The lack of polarization measurements from this region and the orientation of the ``hole'' suggests that the magnetic field should be aligned parallel to the line of sight and would therefore be best detectable using Faraday rotation measurements. Whilst very few such observations exist, the measurements of \citet{wr04} give some hint of increased emission in this regions (suggesting a field which is parallel to the line of sight), though somewhat ambiguously. Thus further observations are necessary to confirm this picture.

\subsection{Relevance to theory}
 One of the primary effects of magnetic fields in the simulations is that, even at field strengths which do not prevent collapse, the field can have a significant influence on the star formation rate in the cloud. For example overall we find that, after 1.5 cloud free-fall times, only 4\% of the gas has been converted into stars for a marginally supercritical collapse ($M/\Phi = 3$) compared with 16\% in the hydrodynamic case. Similar effects of the magnetic field on the star formation rate have been found in MHD calculations of star formation in the presence of driven turbulence \citet{vsetal05}. In part this can be attributed to the simple fact that the magnetic field adds an extra source of pressure support to the cloud. Thus we would expect that in the present context it would be possible to similarly decrease the star formation rate by, for example, simply scaling up the turbulent velocity field or by increasing the temperature of the cloud. However we would also expect the change in the cloud geometry to be rather different under these circumstances, since the magnetic field preferentially supports material perpendicular to field lines. The degree to which this pressure support is anisotropic depends on the ratio of gas-to-magnetic pressure, $\beta$ -- for example a weak field will be much more readily tangled and thus exert a more isotropic pressure, whereas a stronger field will exert a pressure that has a much stronger dependence on the initial field geometry. Furthermore we would not expect to find any of the magnetically driven cloud structures, such as the filaments in the cloud envelope aligned with the global field direction and magnetically supported voids. \citet{pb07} found that the degree to which the magnetic field can be replaced by an equivalent increase in thermal pressure is strongly dependent on the field geometry.


 The net result of the lower star formation rate in the magnetised runs is also a decrease in the importance of violent interactions, leading to fewer ejections and therefore also a trend (though tentative) towards more massive stars being formed (ie. relatively fewer brown dwarfs). Such a trend might have important implications for the variation in the IMF in different environments \citep[e.g.][]{kroupa01} and as a function of galaxy evolution. In the local environment the IMF appears to have a universal shape (at least within the observational uncertainties), although this may be because there is also remarkable uniformity in the inferred level of magnetic field support in observations of local star formation regions \citep{crutcher99} (namely that, as discussed above, most star forming cores appear to be marginally supercritical with super-Alfv\'enic velocity dispersions and $\beta \sim 0.1$).

\subsection{Limitations and future directions}
 Perhaps the most significant limitation of the calculations presented in this paper is that, at the resolution employed, we were not able to study the effect of magnetic fields on the small scale fragmentation of cores, which may be important with respect to the formation of circumstellar discs \citep{pb07} and in affecting fragmentation to form binary systems \citep{machida05b,pb07,ht08,machidaetal07}. A related question which would be interesting to examine is how or whether the global magnetic field affects the frequency and/or strength of jets and outflows in molecular clouds \citep[e.g.][]{mt04,bp06}.

 Whilst increasing the numerical resolution will improve the resolution of the small scale magnetohydrodynamics, there is not strong motivation to do so as the small scale dynamics is also affected by other physical processes which have not been modelled in the present calculations. The most important of these are the effects of radiative transfer (ie. replacing the barytropic equation of state) and physical diffusion in the magnetic field due to non-ideal MHD.
 
 As mentioned in \S\ref{sec:eos}, we expect from preliminary simulations which incorporate a self-consistent treatment of radiative transfer (in the flux-limited diffusion approximation) rather than a barytropic equation of state \citep{wb06} that a full treatment will affect fragmentation where the assumption of spherical symmetry is poor and in the region surrounding a collapsing core where the temperature can increase because of the propagation of radiation from the high temperature central condensation into the lower density gas (a process not captured by a barytropic equation of state where temperature is proportional to density). These issues in particular may be important for fragmentation in discs, which, it should be noted is the prime source of much of the star formation activity in the hydrodynamic and very weak-field ($M/\Phi = 20$) calculations presented here. It is therefore crucial to quantify the degree to which this fragmentation is physical by performing calculations which explicitly evolve the radiation field.
 
  Secondly, whilst flux-freezing is thought to be a good approximation for molecular cloud dynamics on large scales, non-ideal MHD effects including ambipolar (ion-neutral) diffusion \citep[e.g.][]{mp81,sla87}, the effect of finite conductivity and the Hall effect are all important at some level on smaller scales \citep[e.g.][]{wn99}. This in particular applies to the process of core fragmentation (though \citealt{om06} suggest that ambipolar diffusion is unable to set a characteristic mass scale in a turbulent flow because of the continued propagation of compressive slow MHD waves below the ambipolar diffusion scale). It has also been suggested that ambipolar diffusion rates may be enhanced in turbulent flow \citep{heitschetal04,ln04} and therefore that ion-neutral diffusion may be important also in the earlier stages of collapse. Thus it is imperative that the calculations should be extended to include non-ideal MHD effects in order to quantify these effects in the present calculations (for example \citet{hw04a} have already implemented a two-fluid scheme for treating ion-neutral diffusion in an SPH context which could be used in future calculations). Studying physical diffusion processes also requires that the numerical diffusion (e.g. due to the artificial resistivity introduced in order to capture shocks) be reduced to a level below the physical diffusion scale, leading to a much more stringent criterion for fully resolved MHD simulations (of which those presented here are \emph{not}) compared to purely hydrodynamic runs (where it is sufficient to resolve the Jean's length in order to capture the fragmentation scale).


\section*{Acknowledgements} 
We acknowledge useful discussions with Chris Brunt. Thanks also go to Chris McKee for what proved to be a very useful suggestion following a talk on this work. We also thank the referee, Ant Whitworth, for comments which have helped to clarify various points in this paper. DJP is currently supported by a Royal Society University Research Fellowship, although much of this work has been completed whilst funded by a UK PPARC/STFC postdoctoral research fellowship.  MRB is grateful for the support of a Philip Leverhulme Prize and a EURYI Award. This work, conducted as part of the award ``The formation of stars and planets: Radiation hydrodynamical and magnetohydrodynamical simulations"  made under the European Heads of Research Councils and European Science Foundation EURYI (European Young Investigator) Awards scheme, was supported by funds from the Participating Organisations of EURYI and the EC Sixth Framework Programme. Calculations were performed on the United Kingdom Astrophysical Fluids Facility (UKAFF) and Zen, the SGI Altix ICE cluster belonging to the Astrophysics group at the University of Exeter. Visualisations were produced using SPLASH \citep{splashpaper}, a visualisation tool for SPH that is publicly available at http://www.astro.ex.ac.uk/people/dprice/splash. 

\appendix

\bibliography{sph,mhd,starformation}

\label{lastpage}
\enddocument